# Response Style Characterization for Repeated Measures Using the Visual Analogue Scale

Shunsuke Minusa[1], Member, IEEE, Tadayuki Matsumura[1], Kanako Esaki[1], Member, IEEE, Yang Shao[1], Chihiro Yoshimura[1], and Hiroyuki Mizuno[1], Fellow, IEEE

[1] Center for Exploratory Research, Research & Development Group, Hitachi Ltd., Tokyo 185-8601, Japan

Corresponding author: Shunsuke Minusa (e-mail: shunsuke.minusa.hd@hitachi.com).

This work was supported by the Research & Development Group, Hitachi Ltd.

**ABSTRACT** Self-report measures (e.g., Likert scales) are widely used to evaluate subjective health perceptions. Recently, the visual analog scale (VAS), a slider-based scale, has become popular owing to its ability to precisely and easily assess how people feel. These data can be influenced by the response style (RS), a user-dependent systematic tendency that occurs regardless of questionnaire instructions. Despite its importance, especially in between-individual analysis, little attention has been paid to handling the RS in the VAS (denoted as response profile (RP)), as it is mainly used for within-individual monitoring and is less affected by RP. However, VAS measurements often require repeated self-reports of the same questionnaire items, making it difficult to apply conventional methods used for Likert scales. In this study, we developed a novel RP characterization method for various types of repeatedly measured VAS data. This approach involves modeling RP as distributional parameters $\theta$ through a mixture of RS-like distributions, and addressing the issue of unbalanced data through bootstrap sampling for treating repeated measures. We assessed the effectiveness of the proposed method using simulated pseudo-data and an actual dataset from an empirical study. The assessment of parameter recovery showed that our method accurately estimated the RP parameter $\theta$, demonstrating its robustness. Moreover, applying our method to an actual VAS dataset revealed the presence of individual RP heterogeneity, even in repeated VAS measurements, similar to the findings for Likert scales. Our proposed method enables RP heterogeneity-aware VAS data analysis, similar to Likert-scale data analysis.

## I. INTRODUCTION

Monitoring human health is essential in various health-related fields. Both objective and subjective monitoring play an important role in examining human health [1]. Self-reports using specific measurement scales have been widely used to monitor subjective feelings such as the degree of fatigue [2], [3], depression [4], pain [5], and emotions [6]. For example, previous studies assessed patients' pain using various scales including Likert, face, numerical rating, and visual analog scales (VAS) [7], [8].

Recently, the use of VAS has become more popular [7]. The VAS consists of a line whose length is typically 100 mm, with anchor descriptors at both ends of the line (unipolar VAS) or both ends and center (bipolar VAS) [9]. Unlike Likert scales, which require the response to fit descriptors in a categorical or discrete manner (e.g., selection from 5- and 6-point), the VAS enables continuous responses to subjective health conditions, enabling precise assessment of how people feel [10]. Owing to these characteristics, the VAS has been generally accepted as appropriate for comparing conditions and assessing the time course of how the feelings of specific persons change over time (i.e., within-individual assessment). Furthermore, as smartphones become more readily available, researchers can introduce an electronic VAS (e.g., a slider scale) as a substitute for the conventional paper VAS, enabling the digital collection of large amounts of VAS data [11], [12]. This transition has expanded VAS usage not only among within-individual assessments but also among between-individual comparisons.

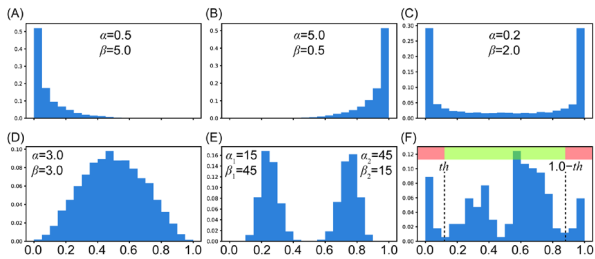

**FIGURE 1.** Examples of response styles (RSs) observed in histograms with the empirical density distribution of VAS data. (A) dis-acquiescent RS, DRS. (B) acquiescent RS, ARS. (C) extreme RS, ERS. (D) midpoint RS, MRS. (E) bimodal MRS, BiMRS. These data were simulated using Beta distribution with specified parameters. (F) Typical histogram of actual VAS responses of some participants.

One of the issues regarding the application of the VAS to between-individual comparisons is the difference in response styles (RSs) among individuals [13], [14]. RS is the systematic tendency not to depend on the scale instructions, and has been studied extensively for Likert scales (Fig. 1). Among well-known RSs for Likert scales, a popular RSs is the disacquiescent RS (DRS, Fig. 1A), in which persons respond to the lower/left end of lines regardless of the instructions, and acquiescent RS (ARS, Fig. 1B), vice versa. The other popular RS is the extreme RS (ERS, Fig. 1C), in which individuals tend to preferentially respond at both ends of the scales. Furthermore, some people, especially Asians, tend to show midpoint RS (MRS, Fig. 1D), in which they prefer to select rating items around the center of the scale [14], [15]. RSs similar to those observed for the Likert scale are also observed for the VAS. We often observed an MRS-like response tendency on both sides across the center (hereafter referred to as bimodal MRS, BiMRS, Fig. 1E) upon using the bipolar VAS as a rating scale, which places the neutral position at the center of the scale, and conflicting concepts at both ends [16]. The VAS has often been used to evaluate relative changes among individuals, such as in clinical studies of pain and fatigue [17], [18]. Consequently, there have been cases in which the absolute values of VAS responses were analyzed across patients [19], [20], with little attention paid to RS-like bias in the VAS. However, these individual variations in RS (i.e., individual RS heterogeneity) may affect inter-individual comparisons of subjective health status [21]. Therefore, RS heterogeneity in the VAS is consequential.

Several RS quantization methods based on Likert scales have been proposed [22]. Previous studies, using simple ideas, represented RSs as the ratio of midpoints or extremes to the total number of questionnaire items [13]. This method is easy to use, but cannot be used to separate the RS from the responses. Other studies attempted to separate RSs and psychological constructs using the anchoring vignette method [23], [24]. In anchoring vignette methods, people report their feelings as well as the status of hypothetical individuals or situations based on short descriptions of the hypothesis (referred to as anchoring vignettes). Using both self-reports and reports on others, the anchoring vignette method separates the RS through statistical modeling. This method can separate RS from responses but requires the modification of questionnaires to carefully design anchoring vignette parts, and satisfy response consistency and vignette equivalence. Similarly, several studies have proposed RS elimination methods without modifying the questionnaires [25]–[28]. Considering RS emerges independently of questionnaire items, it is possible to distinguish instruction-independent RS parameters using item response theory (IRT). Although this third method can eliminate RS without questionnaire modification, it requires that all people answer the items of various questionnaires in the same manner (i.e., every individual rates all questionnaire items on multiple occasions as instructed) to assume task independence. Thus, three types of RS quantization have been proposed for the Likert scale.

Most previous related studies focused on RS handling methods for Likert scales [29]. One reason for this is that most methods rely on IRT modeling for single or multiple questionnaires, including a fixed number of response items [24], [27]. IRT has been vigorously developed from the perspective of test theory, and generally assumes that responses to the same items of specific questionnaires are obtained from all participants only once. With this restriction, IRT-based RS elimination methods estimate the item-independent effects of the questionnaire as RS parameters. The extension of IRT from items with discrete ordinal scales, such as the Likert scale, to that with continuous interval scales including VAS, have already been proposed [30]–[32]. However, repeated VAS measurements, for which the number of responses to items can vary among respondents, and some responses may be omitted, do not satisfy such assumptions. For example, participants #A and #B may have responded to the same scale three and four times, respectively, during the monitoring period. Considering these characteristics, IRT extensions to repeatedly measured VAS are not simple; quantification methods of RS in VAS are required (hereafter referred to as response profile (RP), to distinguish it from RS), especially considering repeated-measures assumptions.

We developed a novel RP characterization method for VAS data. The proposed method models the RP using mixture distributions that represent the RPs using a model selection scheme. We validated its robustness using simulated pseudo-data and evaluated real-world RP characteristics using actual VAS datasets. A preliminary version of this study was reported as a poster presentation at the 45th annual international conference of IEEE Engineering in Medicine and Biology Society (IEEE EMBC 2023) [33].

## II. PROPOSED METHOD

We developed a novel RP characterization method that can be adapted for both VAS and interval scales. The proposed method is composed of (i) response profile modeling using fitting to mixture distributions of known RS-like distributions, and (ii) handling of data imbalance using bootstrap sampling and aggregation. Because the proposed method does not employ IRT modeling, it can be applied to VAS data, including with repeated responses, and different numbers of responses among individuals. Additionally, by adopting the bootstrap method, the proposed method can relieve the effect of item-dependent response tendencies on RP characterization, which is beyond the definition of RP, by adjusting the volume of each questionnaire item.

## A. RESPONSE PROFILE MODELING

In response profile modeling, we estimate the mixture distribution parameters $\theta$, representing the RP of a user *user* (Fig. 2). Here, the proposed method estimates $\theta$ for the user-wise VAS dataset of

$$D_{user} = \{x_1, \ldots, x_n, \ldots, x_N\} \subset (0, 1), \quad (1)$$

where $x$ is the self-reported data obtained using the VAS, normalized to the open interval data between 0 and 1. When $x$ has extreme values (i.e., 0 and 1), the sample $x$ of $D_{user}$ is converted to open interval data using the following [34], [35]:

$$x' = \frac{x(N-1) + 0.5}{N}, \quad (2)$$

where $N$ is the total number of $D_{user}$. The empirical probability density distribution is represented as $Emp(x|D_{user})$.

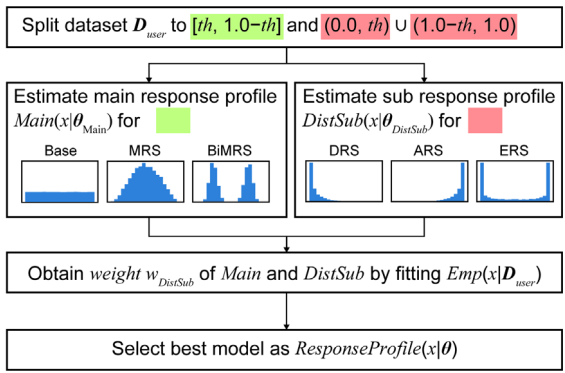

**FIGURE 2.** Schematic diagram of the proposed response profile modeling.

The proposed method approximates $Emp(x|D_{user})$ by the mixture distribution $ResponsePrifile(x|\theta)$. Generally, the probability distributions of unimodal continuous variables are modeled using *Gaussian* and *Beta* [36]–[39]. However, some actual VAS datasets show a non-unimodal distribution, such as a mixture of ERS and BiMRS distributions, as shown in Fig. 1F. Additionally, when we handle RPs in the following analysis, it was desirable to access knowledge of RS for Likert scales. Regarding widely known RSs, ARS/DRS/ERS and MRS are observed at both ends of, and the center of the scale, respectively. Considering these requirements, the proposed method splits dataset $D_{user}$ into sub-datasets $D_{user}^{Main} \subset [th, 1.0-th]$ and $D_{user}^{DistSub} \subset (0.0, th) \cup ()$, and fits each dataset to distributions $Main(x|\theta_{Main})$ and $DistSub(x|\theta_{DistSub})$ using known RSs, and model $RP(x|\theta_{user})$ based on these two distributions, where *th* is the threshold to split the dataset on the dimension of the VAS scale. By connecting RPs with widely known RSs, we can utilize the various types of knowledge reported in RS studies based on a Likert scale.

| **Algorithm 1**: Estimate Main Response Profile | |
|---|---|
| 1 | Construct $D_{user}^{Main}$ by extracting $[th, 1.0-th]$ range from $D_{user} \subset (0.0, 1.0)$ |
| 2 | **for** *DistMain* = {Base, MRS, BiMRS} **do** |
| 3 | Obtain $\theta_{DistMain}$ by fitting $Emp(x|D_{user}^{Main})$ to $DistMain(x|\theta_{DistMain})$ |
| 4 | Calculate fitness indices for $Emp(x|D_{user}^{Main})$ and $DistMain(x|\theta_{DistMain})$ |
| 5 | **end for** |
| 6 | **if** {Bipolar VAS was collected in $D_{user}$} and {Dist(BiMRS($x|\theta_{BiMRS}$) ≥ ACCEPT_BIDIST} and {BiMRS($x|\theta_{BiMRS}$) gets best fitness in *Main*} **do** |
| 7 | Select BiMRS($x|\theta_{BiMRS}$) as Main($x|\theta_{Main}$) |
| 8 | **else if** {MRS($x|\theta_{MRS}$) gets better fitness than Base($x|\theta_{Base}$) } **do** |
| 9 | Select MRS($x|\theta_{MRS}$) as Main($x|\theta_{Main}$) |
| 10 | **else do** |
| 11 | Select Base($x|\theta_{Base}$) as Main($x|\theta_{Main}$) |
| 12 | **end if** |
| 13 | **return** *Main*, $\theta_{Main}$ |

The proposed method estimates $\theta_{Main}$ (Algorithm 1). Here, we assume $Main(x|\theta_{Main})$ can be modeled by MRS($x|\theta_{MRS}$) or BiMRS($x|\theta_{BiMRS}$) if RP exists in $[th, 1.0-th]$. Then, MRS can be represented by an adequate unimodal distribution such as a normal distribution $Gaussian(x|\mu, \sigma)$, where $\mu$ represents the mean, and $\sigma$ represents the standard deviation, or a Beta distribution $Beta(x|\alpha, \beta)$, where $\alpha$ and $\beta$ are shape parameters. Moreover, BiMRS can be modeled using an adequate mixture of two unimodal distributions. For example, when we select $Beta(x|\alpha, \beta)$ as a unimodal distribution, then we select $BetaMixture(x|w_1, \alpha_1, \beta_1, w_2=1-w_1, \alpha_2, \beta_2)$ as a bimodal distribution, where $w$ denotes the weight of $Beta^1(x|\alpha_1, \beta_1)$ and $Beta^2(x|\alpha_2, \beta_2)$.

| **Algorithm 2**: Estimate Sub Response Profiles | |
|---|---|
| 1 | Construct $D_{user}^{DistSub}$ by extracting $(0.0, th) \cup (1.0-th, 1.0)$ range from $D_{user}$ |
| 2 | **for** *DistSub* = {ERS, DRS, ARS} **do** |
| 3 | Obtain $\theta_{DistSub}$ by fitting $Emp(x|D_{user}^{DistSub})$ to $DistSub(x|\theta_{DistSub})$ |
| 4 | Calculate fitness indices for $Emp(x|D_{user}^{DistSub})$ and $DistSub(x|\theta_{DistSub})$ |
| 5 | **end for** |
| 6 | **return** $\theta_{ERS}$, $\theta_{DRS}$, $\theta_{ARS}$ |

We further fit $Main(x|\theta_{Main})$ from $D_{user}^{Main}$ using a model selection scheme with a criterion such Akaike's information criterion (AIC). However, because actual datasets do not always show clear RPs, a simple model selection for a dataset with unclear RPs tends to choose BiMRS, which is more complex in terms of the number of parameters. Thus, in addition to information criterion-based selection, the proposed method also evaluates the degree of separation of the bimodal distribution, which selects the BiMRS with the satisfaction of the smallest information criterion and higher separation than the minimum separation threshold *ACCEPT_BIDIST*. For example, to calculate the degree of separation, we can use a distance $Dist(\theta)$ between peaks of each of the fitted mixture distributions. When we approximate BiMRS by *BetaMixture*, $Dist(\theta)$ can be defined by the difference in the mode of the two Beta distributions, as follows:

$$Dist(\theta_{BiMRS}) = \left| \frac{\alpha_2 - 1}{\alpha_2 + \beta_2 - 2} - \frac{\alpha_1 - 1}{\alpha_1 + \beta_1 - 2} \right| \quad (3)$$

Thus, the proposed method selects $\theta_{Main}$ from $\theta_{MRS}$ or $\theta_{BiMRS}$, or regards no RPs in [$th$, $1.0-th$].

| Algorithm 3: Whole Response Profiles Estimation |
|---|
| 1  $Main$, $\theta_{Main}$ ← Estimate main response profile in Algorithm 1 |
| 2  $\theta_{ERS}$, $\theta_{DRS}$, $\theta_{ARS}$ ← Estimate sub response profiles in Algorithm 2 |
| 3  **for** $DistSub$ = {ERS, DRS, ARS} **do** |
| 4  Obtain $weight$ $w_{DistSub}$ by fitting $Emp(x|D_{user})$ to $w_{DistSub} \times DistSub(x|\theta_{DistSub}) + (1-w_{DistSub}) \times Main(x|\theta_{Main})$ |
| 5  Calculate fitness indices between $Emp(x|D_{user})$ and $w_{DistSub} \times DistSub(x|\theta_{DistSub}) + (1-w_{DistSub}) \times Main(x|\theta_{Main})$ |
| 6  **end for** |
| 7  **If** model better than $Main(x|\theta_{Main})$ exists **then** |
| 8  Select best model components $w_{ADE}$ and $DistSub(x|\theta_{DistSub})$ |
| 9  $ResponseProfile(x|\theta) \leftarrow w_{ADE} \times DistSub(x|\theta_{DistSub}) + (1- w_{ADE}) \times Main(x|\theta_{Main})$ |
| 10  **else** $ResponseProfile(x|\theta) \leftarrow Main(x|\theta_{Main})$ |
| 11  **end if** |
| 12  **return** $ResponseProfile(x|\theta)$ |

The proposed method estimates $\theta_{DistSub}$ based on the RS knowledge (Algorithm 2). Here, we assume $DistSub(x|\theta_{DistSub})$ can be modeled by $ARS(x|\theta_{ARS})$, $DRS(x|\theta_{DRS})$, or $ERS(x|\theta_{ERS})$ if RP exists. Because these RSs show monotonic or U-shaped distributions (Fig. 1A-C), $DistSub(x|\theta_{DistSub})$ is modeled by a Beta distribution $Beta(\alpha_{ADE}, \beta_{ADE})$ satisfying the shape restriction to represent each RS shape, which can flexibly change its distribution shape. The proposed method obtains $\theta_{ARS}$, $\theta_{DRS}$, and $\theta_{ERS}$, which are candidates of $\theta_{DistSub}$, based on $D_{user}^{DistSub}$ to fit to $Beta(\alpha_{ADE}, \beta_{ADE})$, respectively. For each candidate RP, fixing $\theta_{Main}$ and candidates of $\theta_{DistSub}$, the proposed method calculates the optimal mixture weight $w_{DistSub}$, and finally selects the best integration of $\theta_{Main}$, $w_{ADE}$,

and the candidate of $\theta_{DistSub}$ in terms of information criterion for $D_{user}$. For example, when $\theta_{ERS}$ obtains the best fit, the proposed method accepts $\theta_{ERS}$ as $\theta_{DistSub}$.

Finally, the proposed method characterizes RPs as $\theta$ (Algorithm 3). Using the obtained parameters $\theta_{Main}$, $w_{ADE}$, and $\theta_{DistSub}$, we obtain the mixture distribution representing RPs as

$$ResponseProfile(x|\theta) = w_{ADE} \times DistSub(x|\theta_{DistSub}) + (1 - w_{ADE}) \times Main(x|\theta_{Main}) \quad (4)$$

and the RP parameters $\theta$ as $\theta = \{w_{ADE}, \theta_{DistSub}, \theta_{Main}\}$. For example, in the case of Fig. 1F, with a beta distribution representation, the proposed method characterizes ERS and BiMRS as RPs and obtains

$$\theta = \{w_{ADE}, \theta_{ERS}, \theta_{Main}\}$$
$$= \{w_{ADE}, \alpha_{ARS}, \beta_{ARS}, w_1, \alpha_1, \beta_1, w_2 = 1 - w_1, \alpha_2, \beta_2\}. \quad (5)$$

### B. UNBALANCED DATA HANDLING

An unbalanced data handling process adjusts the imbalance of the number of responses for each instruction and evaluates the range of RP parameters $\theta$ (Fig. 3). In contrast to responses to validated questionnaires, in which all users are expected to respond to the same number of questions, the number of responses varies depending on the question in repeated VAS measurements. Because RS and RP are defined as question-independent systematic response tendencies, such an imbalance in the dataset $D_{user}$ may affect bias in the quantification of RPs by item-dependent tendencies. Specifically, for VAS, the tendency to respond to the instruction center emerges as MRS in monopolar VAS scales, and BiMRS in bipolar VAS scales. To relieve and estimate the impact of such bias, the proposed method uses bootstrapping.

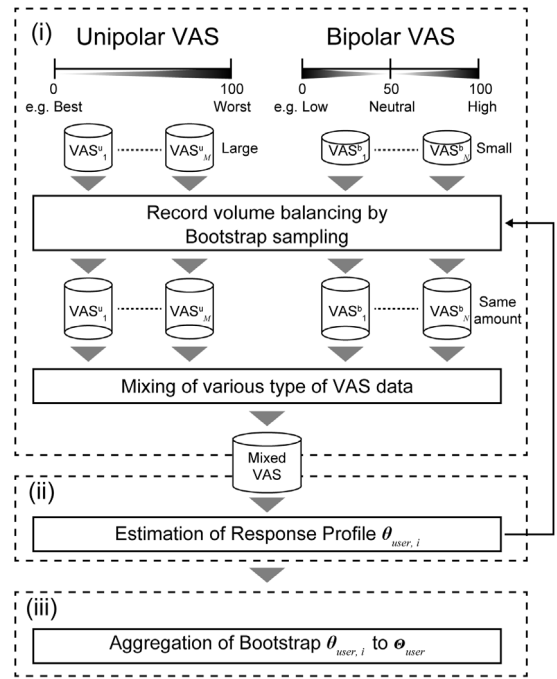

**FIGURE 3.** Schematic diagram of the proposed response profile quantification method. The proposed method comprises (A) Data volume balancing of several types of VAS data, (B) $i$-th response profile $\theta_{user, i}$ estimation, and (C) Bootstrap aggregation process of $\Theta_{user}$ ={$\theta_{user, i}$} to aggregated response profile indicator $\theta_{user}$.

TABLE I
SIMULATION DATA GENERATED BASED ON BETA DISTRIBUTIONS

| # | Model description | $w_1$ | $\alpha_1$ | $\beta_1$ | $w_2$ | $\alpha_2$ | $\beta_2$ | $w_{ADE}$ | $\alpha_{ADE}$ | $\beta_{ADE}$ |
|---|---|---|---|---|---|---|---|---|---|---|
| 11 | ERS | 0.0 | – | – | – | – | – | 1.0 | 0.1 | 0.1 |
| 12 | DRS | 0.0 | – | – | – | – | – | 1.0 | 1.0 | 30.0 |
| 13 | ARS | 0.0 | – | – | – | – | – | 1.0 | 30.0 | 1.0 |
| 14 | MRS-a | 1.0 | 10.0 | 10.0 | – | – | – | – | – | – |
| 15 | MRS-b | 1.0 | 15.0 | 45.0 | – | – | – | – | – | – |
| 16 | MRS-c | 1.0 | 45.0 | 15.0 | – | – | – | – | – | – |
| 17 | BiMRS-a | 0.5 | 15.0 | 45.0 | 0.5 | 45.0 | 15.0 | – | – | – |
| 18 | BiMRS-b | 0.5 | 15.0 | 30.0 | 0.5 | 30.0 | 15.0 | – | – | – |
| 19 | BiMRS-c | 0.5 | 15.0 | 20.0 | 0.5 | 20.0 | 15.0 | – | – | – |
| 21 | ERS-05_MRS-05 | 1.0 | 10.0 | 10.0 | – | – | – | 0.5 | 0.1 | 0.1 |
| 22 | ERS-03_MRS-07 | 1.0 | 10.0 | 10.0 | – | – | – | 0.3 | 0.1 | 0.1 |
| 23 | ERS-01_MRS-09 | 1.0 | 10.0 | 10.0 | – | – | – | 0.1 | 0.1 | 0.1 |
| 24 | DRS-05_MRS-05 | 1.0 | 10.0 | 10.0 | – | – | – | 0.5 | 1.0 | 30.0 |
| 25 | DRS-03_MRS-07 | 1.0 | 10.0 | 10.0 | – | – | – | 0.3 | 1.0 | 30.0 |
| 26 | DRS-01_MRS-09 | 1.0 | 10.0 | 10.0 | – | – | – | 0.1 | 1.0 | 30.0 |
| 31 | ERS-05_BiMRS-05 | 0.5 | 15.0 | 30.0 | 0.5 | 30.0 | 15.0 | 0.5 | 0.1 | 0.1 |
| 32 | ERS-03_BiMRS-07 | 0.5 | 15.0 | 30.0 | 0.5 | 30.0 | 15.0 | 0.3 | 0.1 | 0.1 |
| 33 | ERS-01_BiMRS-09 | 0.5 | 15.0 | 30.0 | 0.5 | 30.0 | 15.0 | 0.1 | 0.1 | 0.1 |
| 34 | DRS-05_BiMRS-05 | 0.5 | 15.0 | 30.0 | 0.5 | 30.0 | 15.0 | 0.5 | 1.0 | 30.0 |
| 35 | DRS-03_BiMRS-07 | 0.5 | 15.0 | 30.0 | 0.5 | 30.0 | 15.0 | 0.3 | 1.0 | 30.0 |
| 36 | DRS-01_BiMRS-09 | 0.5 | 15.0 | 30.0 | 0.5 | 30.0 | 15.0 | 0.1 | 1.0 | 30.0 |

When $D_{user}$ has a data imbalance, the proposed method constructs a sub-dataset of $D_{user}$ by bootstrap sampling. $D_{user}$ comprises $M$ types of unipolar VAS $VAS_m^u$ and $N$ types of bipolar VAS $VAS_n^b$. Each type of VAS may include a different number of responses. The proposed method then conducts sampling with replacement by the type of VAS scale to extract the same number of samples (Fig. 3(i)). When VAS types have a hierarchical structure, such as in the case of both unipolar and bipolar scales, and with different instructions within the scale type, this process can be repeated. After sampling, all the sampled data were used as the dataset for response profile modeling. During sampling, the proposed method constructs an $i$-times sampling of the sub-dataset $D_{user, i}$ from the original dataset $D_{user}$. The sub-dataset $D_{user, i}$ includes the same number of responses by instruction type; therefore, this process can be regarded as the artificial construction of a dataset in which the method can characterize RPs under relieving the effect of item-dependent response tendencies. From the machine learning perspective, this process is also interpreted as class imbalance addressing methods such as oversampling, undersampling, and bagging [40], [41].

Using the sub-dataset $D_{user, i}$, the proposed method evaluates the variability of the RP parameters. After constructing $D_{user, i}$, we obtain $i$ th sampling of RP parameters $\theta$ as $\theta_{user, i}$ by the response profile modeling process (Fig. 3(ii)). However, $\theta_{user, i}$ may differ from the actual RP parameters owing to the biased extraction in the sampling process. To assess such variability, the proposed method repeats this sampling and RP modeling cycles $N$ times as $\Theta_{user} = \{\theta_{user, i}\}$, and estimates the characteristics of this bootstrap samples, including ranges and representative statistics (Fig. 3(iii)). Finally, the proposed method obtains RP parameters $\theta$ by aggregating $\Theta_{user}$.

### III. PSEUDO-DATA SIMULATION

#### A. SETUP OF SIMULATION
To assess the robustness of the proposed method, we performed a parameter recovery experiment using simulated data. We included various RP types and those mixtures with 21 conditions as $\theta_{groundtruth}$ (Table I), regarded as participants with different RPs. For each condition, 1000 samples of response data were extracted from a Beta mixture distribution parameterized by $\theta_{groundtruth}$. The model selection used Akaike's information criterion. Assuming this simulated data satisfies the task independence (i.e., already balanced), we obtained RP parameters $\theta_{user}=\theta_{estim}$ when the hyperparameters were changed; we changed the data splitting threshold $th$ from 0.05 to 0.45 in steps of 0.10, the minimum separation threshold $ACCEPT\_BIDIST$ with three conditions (i.e., 0.00,

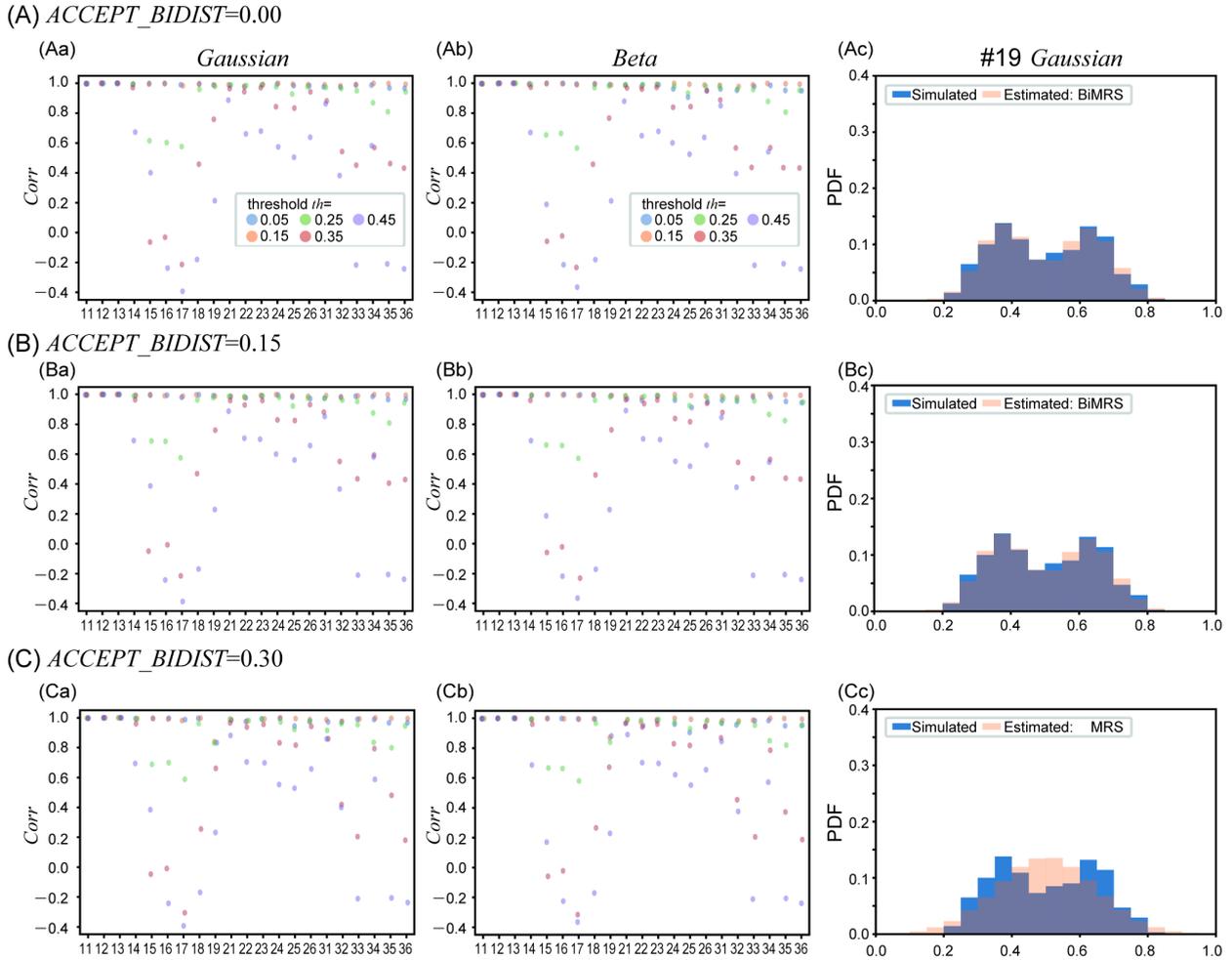

**FIGURE 4.** Parameter recovery results similarity in the shape of the recovered distribution. The similarity is evaluated by Pearson's correlation coefficient, *Corr*. The main response profile distributions are approximated for (A) Gaussian and (B) Beta distributions. Each color depicts the result for hyperparameter *th*. (a)-(c) represent the effect of hyperparameter *ACCEPT_BIDIST*. (C) Empirical and estimated response profile distribution for #19 with *th* = 0.05. As shown in the change from (Cb) to (Cc), the selected *Main*, BiMRS is substituted with MRS.

0.15, 0.30), and unimodal distribution with two conditions (i.e., *Gaussian* and *Beta*).

### B. EVALUATION

We evaluated the robustness of the metrics by changing the hyperparameters. We further calculated Pearson's correlation coefficient for the histograms (*Corr*) to assess how the proposed method approximated the shape of the empirical probability density distribution [42]. For comparison, we obtained empirical probability density distributions using a histogram with a bin width of 0.05. We also analyzed the agreement between the estimated RP parameters $\theta_{estim}$ and $\theta_{groundtruth}$ with both Pearson's correlation coefficient $r$ and its linear regression. Here, $\theta_{groundtruth}$ was defined with *Beta* distributions $Beta(x|\alpha, \beta)$, with mean $\mu$ and standard deviation $\sigma$ calculated as follows:

$$\mu = \frac{\alpha}{\alpha+\beta}, \sigma = \sqrt{\frac{\alpha\beta}{(\alpha+\beta)^2(\alpha+\beta+1)}} \quad (6)$$

Therefore, in the agreement analysis for the *Gaussian* approximation, we compared $w$ and the converted $\mu$ and $\sigma$ based on $\theta_{groundtruth}$, and RP parameters $\theta_{estim}$ estimated with $Gaussian(x|\mu, \sigma)$.

### IV. HUMAN EXPERIMENT

To describe RP characteristics in the real world, we empirically analyzed the RP distribution and the relationship between RPs and participant characteristics using an enhanced version of the experimental dataset called *DailySense*[1] [43]. The experimental design and procedure details have already been reported in [44], from which only the number of participants has been updated. Herein, we briefly describe the experimental information related to this analysis.

---

[1] See: https://dx.doi.org/10.5281/zenodo.10816004.

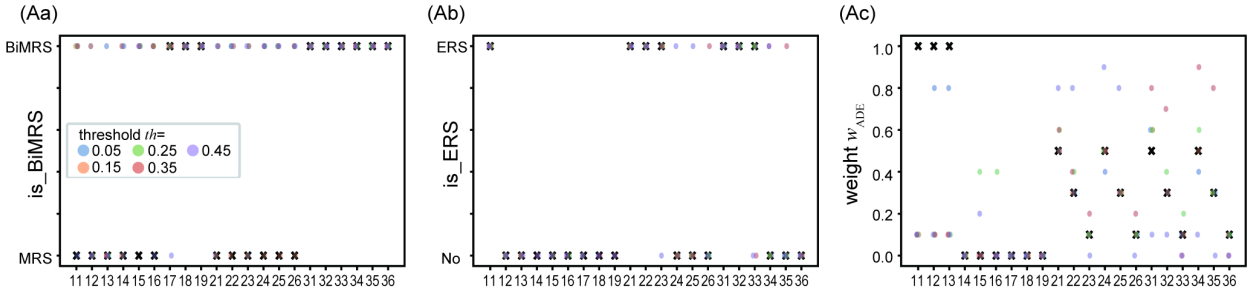

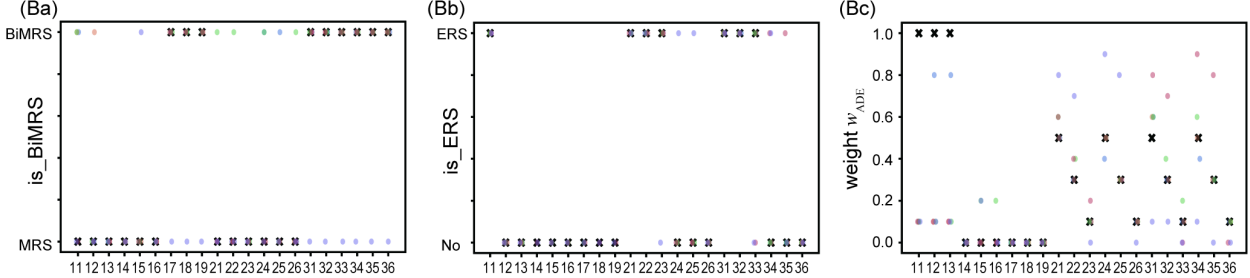

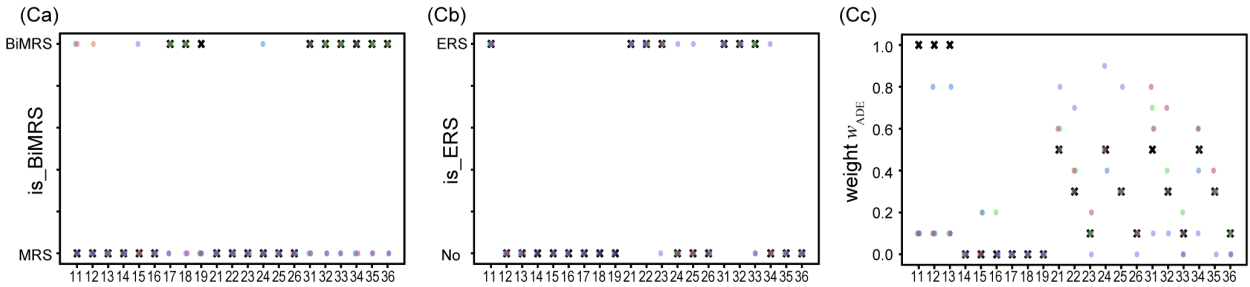

**FIGURE 5.** Parameter recovery results in terms of response profile indicators: (A) *is_BiMRS*, (B) *is_ERS*, and (C) weight of sub-response profile distribution $w_{ADE}$. Each color depicts the result for hyperparameter *th*. (a)-(c) represent the effect of hyperparameter *ACCEPT_BIDIST*. Black crosses represent the groundtruth values. In these results, the main response profile distributions are approximated using a *Gaussian* distribution.

## A. EXPERIMENTAL PROCEDURE

We conducted a consecutive 14-day experiment in a real-life setting composed of a smartphone-based subjective psychological evaluation and physiological sensing. The study protocol was approved by the internal review board of the Research and Development Group, Hitachi, Ltd., and was conducted in accordance with the Declaration of Helsinki. All participants provided informed consent prior to enrollment in the study.

Thirty-six healthy Japanese adult workers who mainly worked remotely (mean±SD, 35.8±7.5; range, 27–58 years; 21 male and 15 female participants) participated in the experiment through two terms. In the first term, 18 participants joined voluntarily, based on their recruitment using company bulletin boards. In the second term, 18 participants were recruited through the participant pool of a research support company, with rewards depending on the engagement rate in the experiments. One participant (#47) was excluded from analysis because they did not complete the personality questionnaire. Accordingly, we analyzed the data of 35 participants (mean±SD, 35.9±7.6; range, 27–58 years; 20 male and 15 female participants).

The participants responded to various VAS and questionnaires using their smartphones. Before and after the 14-day experimental period, participants reported their demographics and completed seven questionnaires, including the Neuroticism Extraversion Openness Five-Factor Inventory (NEO-FFI), to evaluate Big Five personality traits [45]. Participants were randomly notified of the timing of their responses six times per day using an experience sampling method (ESM), also known as ecological momentary assessment (EMA). They reported the two dimensions of emotions as felt in the 30 minutes immediately before response onset. The dimensional emotions (i.e., valence and arousal) were evaluated using the Affective Slider [46], a bipolar VAS scale with face icon instructions on both ends. Participants also reported their degrees of fatigue, stress, anxiety, depression, and sleeplessness using the day reconstruction method (DRM). These five self-reports were evaluated using unipolar VAS with the instruction of "not at all" on the left end as 0 to "worst" on the right end as 1. In both ESM and DRM self-reports, guidelines for the exact midpoint of VAS were not drawn on the scales; that is, participants had to make their responses without an available exact midpoint reference. Finally, we analyzed 2337 records of two types of bipolar VAS in ESM (on average, 66.8±16.4 [times/person]) and 466 records of five types of unipolar VAS in DRM (on average, 13.3±1.2 [times/person]) (Supplementary Table I).

## B. EVALUATION

The proposed method was applied to an actual VAS dataset. Because this dataset was an integration of five and two types of unipolar and bipolar VAS, respectively, and the number of responses varied, we conducted sampling with replacement in two steps to construct $D_{user, i}$. Based on the total number of VAS data, we sampled 300 and 1800 records in the first and second steps, respectively. We experimentally performed the response profile modeling process using *Gaussian* with hyperparameters $th$=0.15 and $ACCEPT\_BIDIST$=0.15. Bootstrap sampling was repeated 1000 times. To simplify the calculation, the mixture weight $w_{ADE}$ was estimated using a 0.1 step resolution (i.e., 0.0 to 1.0 with 0.1 steps).

Note that both valence and arousal were presented as bimodal scales using the Affective Slider [46]. Many studies have treated valence as bimodal, ranging from neutral to either negative or positive, and arousal as unimodal, ranging from low to high. For example, the Self-assessment Manikin (SAM), one of the standard scales for dimensional emotions, treats arousal as unimodal [47]. Conversely, the Affective Slider, which shows a high correlation with SAM, placed two symmetrically mirrored isosceles triangles as an intensity instruction even for arousal, such that arousal could be regarded as bimodal. Our treatment of the arousal scale as bimodal could have led to more unbalanced data collection versus a unimodal treatment. To evaluate the impact on the parameter estimation, we compared both estimation results when arousal scales were treated as bimodal in the bootstrapping process and vice versa.

We evaluated RP characteristics in the real world in two ways. First, we analyzed the stability of the shape of the RP distributions for each bootstrapped $D_{user, i}$. We also checked whether RPs, which may cause analysis bias, emerged in the actual repeated VAS measurements. Additionally, previous studies reported that subjective perception and its output of feelings differ depending on the heterogeneity of individuality [48]–[50]. To clarify whether such RPs are affected by one of the individuality indicators–the participants' personality–we analyzed the relationships between RP parameters and the Big Five personality traits. Here, RP parameters $\theta$ comprised eight distributional parameters, which is shown in Eq. 5 except for $w_2(=1-w_1)$, and four one-shot features (i.e., whether they have BiMRS or MRS, whether they have ERS, whether they have DRS, and whether they have ARS). As Big Five personality indicators, we used standardized scores of five-dimensional personality as assessed by NEO-FFI as $\theta_{dem}$= {$N_t$, $E_t$, $O_t$, $A_t$, $C_t$}, where each represents neuroticism, extraversion, openness, agreeableness, and conscientiousness, respectively (Supplementary Table I). Because we cannot assume normality in the relationship between $\theta$ and $\theta_{dem}$, we evaluated this by Spearman's correlation coefficient $\rho$.

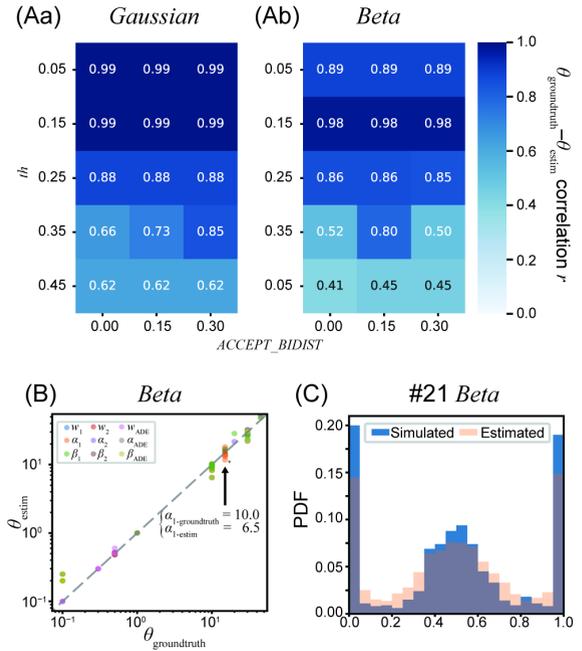

**FIGURE 6.** (A) Parameter agreements in correlation between $\theta_{groundtruth}$ and $\theta_{estim}$ approximated using (a) *Gaussian* and (b) *Beta* distribution. (B) Representative parameter recovery results with $th$=0.15, $ACCEPT\_BIDIST$=0.15, and *Beta*. Parameter agreements between $\theta_{groundtruth}$ and $\theta_{estim}$ are shown. The star* highlights underestimation in $\alpha_L$. (C) Underestimated example in (B). Blue and orange histograms represent the empirical and the estimated distribution of the whole response profile distribution (i.e., MRS+ERS) for simulation #21, respectively.

All data processing and analyses were performed using Python 3.8.13, including SciPy 1.10.1, scikit-learn 1.0, and opencv-python 4.7.0.68. The beta mixture distribution was fitted using a previously proposed method [51]. Statistical significance considered as $p < 0.05$, denoted by an asterisk (*).

## V. RESULTS

We confirmed the performance of the proposed method using simulated data, and analyzed the characteristics of the RP of repeated VAS measurements using actual experimental data.

### A. ASSESSMENT OF RP PARAMETER RECOVERY BY PSEUDO DATA SIMULATION

We confirmed the performance of the proposed method by assessing parameter recovery.

First, we compared the effects of the hyperparameters on the shape of the approximated distribution (Fig. 4). Even when we used *Gaussian* (Fig. 4a) or *Beta* (Fig. 4b) distributions, the proposed method obtained similarly shaped RP distributions based on Corr with varying thresholds $th$ between 0.05 to 0.25. This shape similarity gradually decreased when $th$ was over 0.25. Since larger $th$ (e.g., $th$=0.25) includes both the end and the center ranges of the scales, this indicates robustness in terms of the model distribution and the change in *th* if *DistSub* only affects the end of the VAS scales. However, when the degree of separation of the bimodal distribution was changed, *Corr* changed significantly, especially when both centers of responses were closed (Fig. 4ab, #19). For example, the *Main* RP of #19 was accurately estimated as BiMRS for *ACCEPT_BIDIST*=0.15, and MRS for *ACCEPT_BIDIST*=0.30 (Fig. 4Cbc). This result suggests *ACCEPT_BIDIST* requires appropriate settings based on the data characteristics.

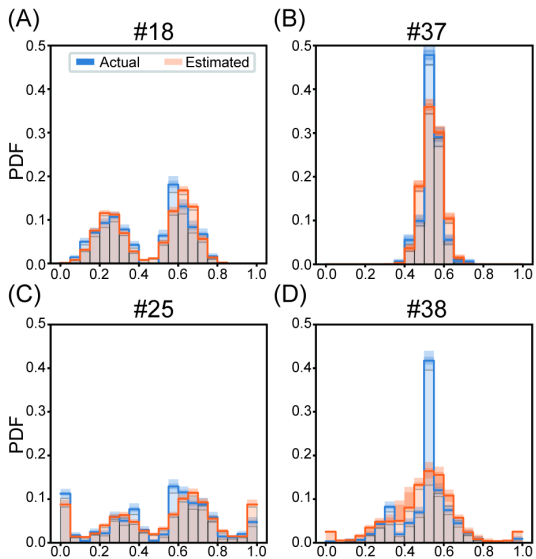

**FIGURE 7.** Response profile distribution by the proposed method for real-world data. (A) BiMRS-like distribution of participant #18, (B) MRS-like distribution of participant #37, (C) Distribution composed of BiMRS-like main RP distribution and ERS-like sub-RP distribution of participant #25, and (D) MRS-like distribution but not fully characterized in terms of its peak concentration of participant #38. Blue and orange histograms represent the empirical distributions of bootstrap samples and estimated RP distributions in each bootstrap trial. Bold lines show the histograms parameterized by the median of bootstrap samples of RP parameters. Boxes around the median lines are 5−95% and 25−75% ranges, respectively.

We further assessed the stability of the estimated parameters, particularly for one of the *Main* RP parameters (*is_BiMRS*), one of the *DistSub* parameters (*is_ERS*), and mixture weight $w_{ADE}$. Because we confirmed the robustness of the estimation results regarding the candidate distribution seed, hereafter, we only show the results of *Gaussian* fitting (Fig. 5), unless otherwise stated. As in the shape similarity analysis, these parameters were correctly estimated even when the hyperparameters were changed with *th* smaller than 0.25, indicating robust estimation of the RP parameters.

Finally, we assessed the agreement of $\theta_{estim}$ and $\theta_{groundtruth}$. The analysis showed that the agreement was less affected by the approximated distribution (i.e., *Gaussian* and *Beta*) and *ACCEPT_BIDIST*, but changed depending on *th* (Fig. 6A). The agreement became best when the threshold *th* approximated 0.15, although the agreement gradually decreased when *th* became large (see Supplementary Table II for details). This result indicates that data thresholding should be adequately considered. For more comprehensive analysis, we focused on the better agreement case (i.e., with *Beta* mixture modeling, *th*=0.15, *ACCEPT_BIDIST*=0.15; Pearson's correlation coefficient, $r$=0.99, $p$<0.001; $y$=1.01$x$−0.19, $R^2$=0.97; Fig. 6B). This result showed good linear relationships, but several parameters, such as $\alpha_1$, seemed to be slightly underestimated. For a detailed analysis, we confirmed the results in #21 as the worst underestimation case (Fig. 6C). When we compared the shape of the RP distribution, the height of ERS and the estimated peak showed a slight difference. However, most of the characteristics of the ERS and unimodality were sufficiently approximated by the proposed method. These results indicate the acceptable RP characterization using the proposed method, unless several parameters are underestimated.

### B. RP CHARACTERIZATION FOR EMPIRICAL STUDY
Using the proposed method, we evaluated the emergence of RP in an empirical study using repeated VAS measurements. Fig. 7 shows the representative types of RP characterizations. Through unbalanced data handling, the proposed method estimated the range of RP depending on each questionnaire response; for example, in #25, $\mu_1$, the mean of the first peak, was estimated as 0.66 (0.30–0.67; 95% bootstrap confidence interval, CI). Although the ranges of estimated parameters may be slightly wider, similar RP distribution results were observed based on the estimated parameters (Fig. 7C). Even if the arousal scale was treated as unimodal, most of the ranges of estimated parameters overlapped (e.g., 0.30–0.67 for the case treated in bimodal manner vs. 0.31–0.67 for the case treated in unimodal manner, $\mu_1$ for #25; see Supplementary Table III and IV for details). These results suggest the robustness of our method by introducing unbalanced data handling. This indicates that RP in the VAS stably emerges, similar to studies using Likert scales. Hereafter, we only show the analysis results in which the arousal scale was treated as bimodal in the unbalanced data handling process.

We comprehensively analyzed each characterization result. Even in a study using both bipolar and unipolar VASs, although most participants (34 of 35) showed BiMRS-like tendencies as *Main* RP, the proposed method clarified the existence of both BiMRS-like and MRS-like RPs (Fig. 7AB). Moreover, some participants had both *Main* and *DistSub* RPs, such as the BiMRS and ERS (Fig. 7C). Regarding *DistSub* RPs, our method quantified 14 participants with an ERS tendency and three with a DRS tendency, but none with an ARS tendency. Despite this success in RP characterization, a few participants showed insufficient fit. For example, for participant #38, the proposed method cannot fully approximate the shape of the response distribution in terms of its peak concentration, which seems to be a limitation of using a *Gaussian* distribution (Fig. 7D). In this case, the proposed method also identified #38 as BiMRS-like RP with larger weight $w_1$=0.83 (CI, 0.16–1.00) but it might appear MRS-like RP based on visual inspection. This tendency was also observed when the CI of $w_1$ included 1.00 in its range (3 out of 35).

Finally, we analyzed the relationships between the RP parameters and the participants' personalities. Regarding background demographics, the participants exhibited five personalities, as shown in Table II. The correlation analysis indicated almost no clear relationship between RP parameters and Big-Five parameters (Table III), except for the mean of the left peak $\mu_L$ and conscientiousness ($\rho = -0.39$, $p=0.020$). This result indicates that RP parameters may be unique characteristics as distinct from the respondents' personalities.

## VI. DISCUSSION

We developed a novel method to characterize RSs in VAS measurements (i.e., RPs) consisting of response profile modeling and unbalanced data handling. We further assessed the robustness of the proposed method using a parameter recovery experiment with simulated pseudo data. The proposed method was also applied to evaluate the RP characteristics of an empirical study dataset, clarifying the existence of RP heterogeneity, even in repeated VAS measurements.

We confirmed the characteristics of the proposed method through simulated and actual data analyses. The proposed method has three hyperparameters: degree of separation of the two peaks *ACCEPT_BIDIST*, data-split threshold *th*, and types of candidate distributions. We showed that *ACCEPT_BIDIST* might change the characterization of *Main* to MRS or BiMRS (Fig. 4C). *ACCEPT_BIDIST* is a parameter that provides the constraints for selecting a complex bimodal model. Considering its background, this characteristic does not cause severe issues when only one *Main* RP exists (e.g., when the applied dataset is constructed only from unipolar or bipolar VAS responses). However, other cases using both scales, such as the human experiment in this study, require careful setting of the peak separation parameter. In particular, the proposed method sometimes selected BiMRS-like RP different from the visual inspection-based MRS-like envelope in the model selection process, although the bootstrap CI implied the possibility of MRS-like RP (Fig. 7D). For more stable RP characterization, in future studies we must consider better model selection criteria to distinguish whether RP has MRS or BiMRS.

Moreover, in some cases, the proposed method showed a slightly lower approximation of the RP distribution than that of the empirical existence distribution (Figs. 6C and 7D). This is partly because of the characteristics of the candidate distributions. *Gaussian* and *Beta* distributions are the first choices to represent a unimodal probability distribution. These distributions are limited to specific shape characteristics, especially in the center concentrations, and the length of the tail. Although an improvement in fit is not necessarily needed when only RP classification is required because the proposed method sufficiently captures the RP-related shape characteristics, better fit is desirable when RP parameters themselves are used in the following between-individual comparison, such as for controlling covariates. Regarding the use of these RP parameters, introducing more parameter distributions, such as the Student's *t* distribution and Laplace distribution, may improve model fit [52]. Another possible reason for inadequate fit may be the conversion of the dataset using Eq. 2. This conversion was used to omit extreme values and may have particularly affected approximating ERS-like RS as *DistSub*. Updates for better ERS approximation will be a topic of our future studies.

The empirical analysis showed heterogeneity of actual RP emergence in VAS measurements. In our analysis, most participants identified their MRS or BiMRS tendency as *Main*

TABLE II
CORRELATIONS AMONG BIG FIVE PERSONALITY TRAITS

|  | $N_t$ | $E_t$ | $O_t$ | $A_t$ | $C_t$ |
|---|---|---|---|---|---|
| Neuroticism | — | **–0.38**[*] | 0.09 | 0.01 | –0.33 |
| Extraversion | — | — | –0.04 | 0.21 | 0.13 |
| Openness | — | — | — | –0.08 | **0.36**[*] |
| Agreeableness | — | — | — | — | 0.02 |
| Conscientiousness | — | — | — | — | — |

$N_t$, standardized *t*-score of Neuroticism; $E_t$, Extraversion; $O_t$, Openness; $A_t$, Agreeableness; and $C_t$, Conscientiousness; *$p < 0.05$.

TABLE III
CORRELATIONS BETWEEN BIG FIVE PERSONALITY TRAITS AND RP PARAMETERS

|  | $w_1$ | $\mu_1$ | $\sigma_1$ | $\mu_2$ | $\sigma_2$ | $w_{ADE}$ | $\alpha_{ADE}$ | $\beta_{ADE}$ | *is_BiMRS* | *is_ERS* | *is_DRS* |
|---|---|---|---|---|---|---|---|---|---|---|---|
| Neuroticism | 0.05 | 0.18 | –0.22 | 0.18 | –0.08 | –0.31 | –0.07 | –0.07 | –0.25 | –0.30 | –0.03 |
| Extraversion | 0.13 | –0.17 | 0.06 | –0.20 | 0.16 | 0.27 | 0.21 | 0.23 | –0.07 | 0.29 | 0.05 |
| Openness | –0.11 | –0.14 | –0.01 | 0.13 | 0.02 | 0.01 | 0.07 | 0.07 | 0.05 | 0.01 | –0.06 |
| Agreeableness | 0.24 | 0.25 | –0.07 | –0.13 | –0.13 | –0.13 | –0.04 | –0.04 | 0.19 | –0.13 | –0.02 |
| Conscientiousness | –0.11 | **–0.39**[*] | 0.26 | 0.14 | 0.10 | 0.18 | -0.02 | 0.00 | 0.02 | 0.10 | 0.07 |

Note that the results of *is_ARS* are not shown as no participant exhibited *is_ARS*. *$p < 0.05$.

RP, although the intensity of how they preferred midpoint selection differed among participants (Supplementary Table I). When we interpret BiMRS as a midpoint preference between neutral and the instructions at both ends, this result is consistent with previous studies that showed a clear MRS tendency in Japanese respondents [14], [15]. This heterogeneity in the modality also follows previous studies, in which various emotion measurement datasets using VAS showed bimodality between 20.3–48.9%, regardless of whether the location guidelines of the scales were initially set to the exact midpoint or not [16]. Since midpoint guidelines were not shown on the VAS scales in this experiment, this tendency was likely not a consequence of our data collection procedures, but rather represented participants' characteristics. This result may imply the effectiveness of *ACCEPT_BIDIST*=0.15, which we experimentally set for this dataset, for other general datasets handling self-reports. In addition to *Main* RP, *DistSub* RP varied among participants. Unlike the conventional VAS usage of within-individual comparisons, RP heterogeneity can distort the results of between-individual VAS comparisons, similar to cross-sectional studies using Likert scales [13], [53]. Moreover, the RP parameters had fewer relationships with the participants' personalities, indicating the difficulty in controlling the effect by introducing demographic covariates (Table III). Our analysis suggests that we must consider the existence of RP heterogeneity, which can degrade inter-individual analysis, even for real-world VAS measurements.

Finally, this study has limitations. In contrast to RS removal methods [25]–[27], it is challenging to extend our method directly to RP removal methods. Based on the realistic constraints that users rated various self-report questionnaires with repeated measures, our method assumes that users repeatedly respond to various questionnaire items a different number of times. Owing to this constraint, we cannot simply utilize the task independence of the previously proposed methods, which assume that various users answer fixed questionnaire items the same number of times, especially once per item. Although RP characterization is helpful in terms of using RP parameters as covariates, there is need for future studies to extend this to enable RP removal. Moreover, we applied the proposed method to a single empirical dataset. This *DailySense* dataset was limited to a small Japanese population [43], [44]. To confirm the deleterious effects of RP heterogeneity, the proposed method should be utilized with larger datasets that include participants with diverse cultural backgrounds, namely cross-cultural studies.

## VII. CONCLUSION

This study proposes a novel characterization method for the RSs observed in the VAS (denoted as the response profile, RP). Our proposed method robustly identified RPs in simulated and repeatedly measured VAS datasets. These findings suggest the existence of participant-dependent RP heterogeneity, which can degrade inter-individual analyses. These results elucidate the importance of RP evaluation, even in the analysis of self-reported data obtained by VAS. Future RP-aware analyses of repeatedly measured VAS data will offer the possibility of RP heterogeneity-less inter-individual analyses, enabling more precise subjective health research.

## ACKNOWLEDGMENT
The authors thank Dr. Hirohiko Kuratsune and Sakura Tatsumi of Osaka Metropolitan University for their helpful suggestions regarding the human experiments. The authors also thank ASMARQ Co., Ltd. for recruiting participants and supporting procedural operations during the second term of the human experiment.

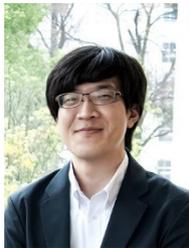

**SHUNSUKE MINUSA** (M'24) was born in Japan. He received his B.E. degree and Master's degree in Information Science and Technology from Hokkaido University, Sapporo, Japan, in 2016 and 2018, respectively.

Since 2018, he has been a researcher at the Hitachi, Ltd. Research & Development Group. His current research interests include human sensing, biostatistics, and psychophysiological data analysis for healthcare applications.

He is a member of the Behaviormetric Society of Japan and the Japan Ergonomics Society.

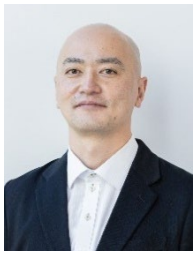

**TADAYUKI MATSUMURA** was born in Japan. He received his B.S. and M.S. degrees in Computer Science from Kyushu University in 2007 and 2009, respectively. Since 2009, he has been a researcher at the Hitachi, Ltd. Research & Development Group. His current research interests include artificial multi-agent system, AI ethics, and emotional intelligence for social AI/Robot applications.

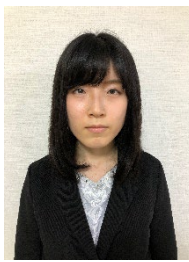

**KANAKO ESAKI** (M'16) was born in Japan. She received the B.S. degree, the M.S. degree, and the Ph.D. degree in mechanical engineering from Waseda University, Tokyo, Japan, in 2010, 2012, and 2024, respectively.

In 2012, she joined the Central Research Laboratory, Hitachi, Ltd., Tokyo., and she has been mainly involved in research into robotics and automation systems. Her current research interests include bio-inspired information processing systems.

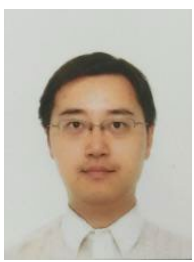

**YANG SHAO** was born in China. He received his B.S. degree in mathematics and physics from Tsinghua University, Beijing, in 2007 and his Ph.D. in numerical analysis from the University of Tokyo, Japan, in 2013.

Since 2013, he has been a researcher at the Hitachi, Ltd. Research & Development Group. His current research interests include large language models, reinforcement learning and artificial intelligence.

He is a member of the Japanese Society for Artificial Intelligence

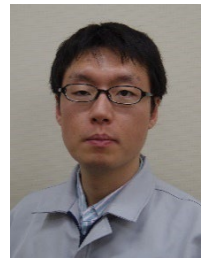

**CHIHIRO YOSHIMURA** was born in Japan. He received the B.A. degree in environmental information and the M.A. degree in media and governance from Keio University, Kanagawa, Japan, in 2005 and 2007, respectively.

In 2007, he joined the Central Research Laboratory, Hitachi Ltd., Tokyo, Japan, where he has been engaged in the research and development of supercomputers and mission critical servers, including vector processors, interconnect switches, and network controllers. Since 2012, he has been engaged in the research of new-paradigm computing including quantum computing and domain specific computing. Currently, he is a Manager and Chief Researcher in Center for Exploratory Research, Research and Development Group, Hitachi Ltd.

He is a member of the Information Processing Society of Japan.

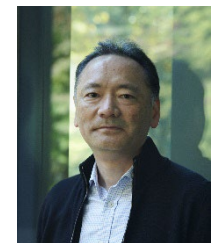

**HIROYUKI MIZUNO** (M'93–SM'17–F'22) received the M.S., and Dr. Eng. degrees in electronic engineering from Osaka University, Osaka, Japan, in 1993, and 2001, respectively. In 1993 he joined Hitachi, Ltd., Tokyo, Japan where he has been engaged in the research and development of high-speed and low-power semiconductor circuits for SRAMs, microprocessors (SuperH), system-on-a-chip (SH-Mobile). He is a pioneer in quantum-inspired computers (CMOS Annealing). From 2002 to 2003 he was a visiting scholar at the Department of Computer Science, Stanford University, Stanford, CA.

He is currently a distinguished researcher in the Center for Exploratory Research (CER), and the laboratory manager of the Hitachi-Kyoto University Laboratory at Hitachi. His current research interests include quantum computing, artificial-intelligence systems using cognitive neuroscientific methods, and Cyber-Human Social System (CHSS). He has served in various roles for international conferences, including the technical program committees of the IEEE International Solid-State Circuits Conference (ISSCC) from 2004 to 2007. He is a Fellow of the IEEE, and a member of ACM.

SUPPLEMENTARY TABLE I
Participants' information and RP one-hot features

| # | gender | Number of Responses | | Personality | | | | | Estimated RP one-hot features | | | | |
|---|---|---|---|---|---|---|---|---|---|---|---|---|---|
| | | Answered ESM records | Answered DRM | Neuroticism | Extraversion | Openness | Agreeableness | Conscientiousness | is_MRS | is_BiMRS | is_ERS | is_DRS | is_ARS |
| 11 | F | 66 | 14 | 54.8 | 40.5 | 51.1 | 57.2 | 60.9 | 0 | 1 | 0 | 0 | 0 |
| 12 | M | 42 | 13 | 46.2 | 57.6 | 51.1 | 54.0 | 44.3 | 0 | 1 | 1 | 0 | 0 |
| 13 | M | 56 | 13 | 53.0 | 43.1 | 64.3 | 34.9 | 55.4 | 0 | 1 | 1 | 0 | 0 |
| 14 | M | 76 | 14 | 63.8 | 50.4 | 43.6 | 56.2 | 29.4 | 0 | 1 | 0 | 0 | 0 |
| 15 | F | 80 | 13 | 56.4 | 55.5 | 66.8 | 57.2 | 60.9 | 0 | 1 | 0 | 0 | 0 |
| 16 | F | 77 | 14 | 67.3 | 45.5 | 45.8 | 44.5 | 37.3 | 0 | 1 | 0 | 0 | 0 |
| 17 | M | 28 | 11 | 42.2 | 41.3 | 47.4 | 58.3 | 57.2 | 0 | 1 | 0 | 1 | 0 |
| 18 | M | 15 | 14 | 43.5 | 54.0 | 56.8 | 47.7 | 48.0 | 0 | 1 | 0 | 0 | 0 |
| 19 | M | 70 | 13 | 53.0 | 55.8 | 49.2 | 56.2 | 55.4 | 0 | 1 | 0 | 0 | 0 |
| 21 | M | 79 | 14 | 71.9 | 34.0 | 60.6 | 54.0 | 33.1 | 0 | 1 | 0 | 0 | 0 |
| 22 | M | 52 | 14 | 53.0 | 63.1 | 60.6 | 43.4 | 53.5 | 0 | 1 | 1 | 0 | 0 |
| 23 | M | 59 | 14 | 57.0 | 41.3 | 66.2 | 64.7 | 59.1 | 0 | 1 | 0 | 0 | 0 |
| 24 | F | 77 | 14 | 72.0 | 37.2 | 59.8 | 48.7 | 30.0 | 0 | 1 | 0 | 0 | 0 |
| 25 | F | 57 | 13 | 58.0 | 50.5 | 49.3 | 63.6 | 28.2 | 0 | 1 | 1 | 0 | 0 |
| 26 | M | 70 | 13 | 48.9 | 43.1 | 26.6 | 56.2 | 33.1 | 0 | 1 | 0 | 0 | 0 |
| 27 | F | 40 | 10 | 50.2 | 52.2 | 52.8 | 36.0 | 35.5 | 0 | 1 | 0 | 1 | 0 |
| 28 | M | 50 | 10 | 48.9 | 43.1 | 45.5 | 43.4 | 55.4 | 0 | 1 | 0 | 0 | 0 |
| 29 | M | 67 | 14 | 51.6 | 26.7 | 73.8 | 43.4 | 68.3 | 0 | 1 | 0 | 0 | 0 |
| 31 | F | 78 | 14 | 51.7 | 38.8 | 52.8 | 36.0 | 46.4 | 0 | 1 | 1 | 0 | 0 |
| 32 | F | 67 | 12 | 68.9 | 37.2 | 68.6 | 50.9 | 44.5 | 0 | 1 | 0 | 0 | 0 |
| 33 | M | 57 | 14 | 31.4 | 57.6 | 54.9 | 58.3 | 55.4 | 0 | 1 | 1 | 0 | 0 |
| 34 | F | 79 | 14 | 64.2 | 33.8 | 59.8 | 25.3 | 53.6 | 0 | 1 | 0 | 0 | 0 |
| 35 | F | 66 | 14 | 45.5 | 45.5 | 52.8 | 50.9 | 50.0 | 0 | 1 | 0 | 0 | 0 |
| 36 | M | 60 | 10 | 35.4 | 46.7 | 66.2 | 10.0 | 77.6 | 0 | 1 | 1 | 0 | 0 |
| 37 | F | 69 | 14 | 72.0 | 48.8 | 52.8 | 38.1 | 51.8 | 1 | 0 | 0 | 0 | 0 |
| 38 | M | 80 | 14 | 51.6 | 48.5 | 47.4 | 47.7 | 53.5 | 0 | 1 | 1 | 0 | 0 |
| 39 | M | 83 | 14 | 47.6 | 66.7 | 54.9 | 51.9 | 64.6 | 0 | 1 | 0 | 0 | 0 |
| 41 | F | 80 | 14 | 56.4 | 42.2 | 40.5 | 29.6 | 40.9 | 0 | 1 | 1 | 0 | 0 |
| 42 | M | 81 | 14 | 42.2 | 63.1 | 53.0 | 51.9 | 49.8 | 0 | 1 | 1 | 0 | 0 |
| 43 | M | 74 | 13 | 61.1 | 46.7 | 49.2 | 58.3 | 83.1 | 0 | 1 | 1 | 0 | 0 |
| 44 | F | 83 | 14 | 73.6 | 50.5 | 61.6 | 48.7 | 64.5 | 0 | 1 | 0 | 1 | 0 |
| 45 | F | 81 | 14 | 86.1 | 45.5 | 65.1 | 53.0 | 48.2 | 0 | 1 | 1 | 0 | 0 |
| 46 | F | 75 | 14 | 48.6 | 47.2 | 52.8 | 40.2 | 64.5 | 0 | 1 | 1 | 0 | 0 |
| 48 | M | 79 | 14 | 62.4 | 48.5 | 54.9 | 60.4 | 49.8 | 0 | 1 | 0 | 0 | 0 |
| 49 | M | 84 | 14 | 21.9 | 63.1 | 68.1 | 56.2 | 72.0 | 0 | 1 | 1 | 0 | 0 |

M, Male; F, Female; For each ESM record, participants reported their two dimensional emotions.
For each DRM record, participant reported their five subjective health conditions.

SUPPLEMENTARY TABLE II
Parameter agreement analysis

| Distribution | th | ACCEPT_BIDIST | Pearson's correlation | | Linear regression | | |
|---|---|---|---|---|---|---|---|
| | | | r | p value | Coefficient | Intercept | $R^2$ |
| Gaussian | 0.05 | 0.00 | 0.99 | <0.001 | 1.49 | -0.15 | 0.98 |
| Gaussian | 0.05 | 0.15 | 0.99 | <0.001 | 1.50 | -0.15 | 0.98 |
| Gaussian | 0.05 | 0.30 | 0.99 | <0.001 | 1.50 | -0.15 | 0.98 |
| Gaussian | 0.15 | 0.00 | 0.99 | <0.001 | 0.95 | 0.03 | 0.98 |
| Gaussian | 0.15 | 0.15 | 0.99 | <0.001 | 0.95 | 0.03 | 0.98 |
| Gaussian | 0.15 | 0.30 | 0.99 | <0.001 | 0.95 | 0.03 | 0.98 |
| Gaussian | 0.25 | 0.00 | 0.89 | <0.001 | 0.54 | 0.19 | 0.79 |
| Gaussian | 0.25 | 0.15 | 0.89 | <0.001 | 0.54 | 0.18 | 0.79 |
| Gaussian | 0.25 | 0.30 | 0.89 | <0.001 | 0.54 | 0.19 | 0.79 |
| Gaussian | 0.35 | 0.00 | 0.67 | <0.001 | 0.38 | 0.24 | 0.45 |
| Gaussian | 0.35 | 0.15 | 0.73 | <0.001 | 0.42 | 0.22 | 0.54 |
| Gaussian | 0.35 | 0.30 | 0.86 | <0.001 | 0.51 | 0.22 | 0.74 |
| Gaussian | 0.45 | 0.00 | 0.63 | <0.001 | 0.40 | 0.21 | 0.39 |
| Gaussian | 0.45 | 0.15 | 0.63 | <0.001 | 0.40 | 0.22 | 0.39 |
| Gaussian | 0.45 | 0.30 | 0.63 | <0.001 | 0.40 | 0.22 | 0.39 |
| Beta | 0.05 | 0.00 | 0.90 | <0.001 | 1.02 | -0.42 | 0.81 |
| Beta | 0.05 | 0.15 | 0.90 | <0.001 | 1.01 | -0.67 | 0.80 |
| Beta | 0.05 | 0.30 | 0.89 | <0.001 | 1.00 | -0.78 | 0.79 |
| Beta | 0.15 | 0.00 | 0.98 | <0.001 | 1.01 | -0.08 | 0.97 |
| Beta | 0.15 | 0.15 | 0.99 | <0.001 | 1.01 | -0.19 | 0.97 |
| Beta | 0.15 | 0.30 | 0.98 | <0.001 | 1.00 | -0.29 | 0.96 |
| Beta | 0.25 | 0.00 | 0.86 | <0.001 | 1.59 | -0.15 | 0.74 |
| Beta | 0.25 | 0.15 | 0.86 | <0.001 | 1.58 | -0.80 | 0.75 |
| Beta | 0.25 | 0.30 | 0.86 | <0.001 | 1.56 | -1.13 | 0.73 |
| Beta | 0.35 | 0.00 | 0.53 | <0.001 | 3.81 | 12.52 | 0.28 |
| Beta | 0.35 | 0.15 | 0.81 | <0.001 | 3.75 | -0.99 | 0.65 |
| Beta | 0.35 | 0.30 | 0.51 | <0.001 | 1.72 | -0.81 | 0.26 |
| Beta | 0.45 | 0.00 | 0.41 | <0.001 | 21.75 | 142.34 | 0.17 |
| Beta | 0.45 | 0.15 | 0.46 | <0.001 | 2.55 | 17.08 | 0.21 |
| Beta | 0.45 | 0.30 | 0.46 | <0.001 | 2.55 | 17.08 | 0.21 |

SUPPLEMENTARY TABLE III
Estimated RP Parameters and Their Evaluation Metrics When Arousal Scale Was Treated As Bimodal

| # | $w_1$ | $\mu_1$ | $\sigma_1$ | $\mu_2$ | $\sigma_2$ | $w_{ADE}$ | $\alpha_{ADE}$ | $\beta_{ADE}$ | D_KL | Corr | ChiSq | Intersect | Bhattacharyya |
|---|---|---|---|---|---|---|---|---|---|---|---|---|---|
| 11 | 0.50 (0.44–0.56) | 0.62 (0.38–0.63) | 0.08 (0.07–0.09) | 0.39 (0.38–0.63) | 0.08 (0.07–0.09) | 0.00 (0.00–0.00) | – | – | 0.14 (0.11–0.19) | 0.85 (0.82–0.88) | 0.34 (0.24–0.49) | 0.79 (0.76–0.82) | 0.18 (0.16–0.21) |
| 12 | 0.73 (0.23–0.78) | 0.71 (0.29–0.73) | 0.10 (0.08–0.11) | 0.32 (0.29–0.73) | 0.10 (0.07–0.11) | 0.20 (0.20–0.20) | 0.20 (0.20–0.25) | 0.20 (0.20–0.25) | 0.11 (0.09–0.14) | 0.79 (0.72–0.84) | 0.43 (0.26–0.86) | 0.83 (0.81–0.85) | 0.19 (0.17–0.21) |
| 13 | 0.58 (0.37–0.63) | 0.42 (0.40–0.68) | 0.09 (0.09–0.10) | 0.66 (0.41–0.68) | 0.09 (0.09–0.10) | 0.30 (0.30–0.30) | 0.15 (0.15–0.15) | 0.15 (0.15–0.15) | 0.12 (0.10–0.14) | 0.79 (0.75–0.83) | 0.38 (0.26–0.59) | 0.82 (0.80–0.84) | 0.19 (0.18–0.21) |
| 14 | 0.62 (0.35–0.65) | 0.66 (0.32–0.67) | 0.06 (0.05–0.07) | 0.33 (0.32–0.67) | 0.07 (0.05–0.08) | 0.00 (0.00–0.00) | – | – | 0.06 (0.05–0.09) | 0.94 (0.91–0.95) | 0.18 (0.11–0.53) | 0.86 (0.83–0.88) | 0.13 (0.11–0.15) |
| 15 | 0.52 (0.41–0.59) | 0.60 (0.34–0.62) | 0.10 (0.09–0.10) | 0.36 (0.34–0.62) | 0.10 (0.09–0.10) | 0.00 (0.00–0.00) | – | – | 0.04 (0.03–0.05) | 0.95 (0.93–0.97) | 0.04 (0.02–0.08) | 0.91 (0.88–0.93) | 0.12 (0.11–0.14) |
| 16 | 0.55 (0.41–0.60) | 0.63 (0.40–0.64) | 0.09 (0.08–0.10) | 0.41 (0.40–0.64) | 0.09 (0.08–0.10) | 0.00 (0.00–0.00) | – | – | 0.13 (0.10–0.16) | 0.84 (0.80–0.87) | 0.35 (0.25–0.50) | 0.79 (0.76–0.81) | 0.18 (0.16–0.20) |
| 17 | 0.69 (0.20–0.81) | 0.62 (0.34–0.64) | 0.10 (0.09–0.13) | 0.40 (0.33–0.64) | 0.10 (0.09–0.13) | 0.10 (0.10–0.10) | 1.00 (1.00–1.00) | 19.0 (18.0–20.0) | 0.14 (0.12–0.17) | 0.80 (0.76–0.84) | 0.23 (0.16–0.33) | 0.78 (0.75–0.81) | 0.21 (0.19–0.23) |
| 18 | 0.54 (0.44–0.56) | 0.62 (0.24–0.63) | 0.07 (0.06–0.08) | 0.25 (0.24–0.63) | 0.07 (0.06–0.08) | 0.00 (0.00–0.00) | – | – | 0.07 (0.05–0.09) | 0.91 (0.87–0.94) | 0.09 (0.06–0.13) | 0.85 (0.83–0.88) | 0.15 (0.13–0.16) |
| 19 | 0.62 (0.35–0.65) | 0.37 (0.36–0.62) | 0.06 (0.04–0.06) | 0.62 (0.36–0.62) | 0.04 (0.04–0.06) | 0.00 (0.00–0.00) | – | – | 0.14 (0.10–0.29) | 0.93 (0.91–0.95) | 0.61 (0.29–2.70) | 0.82 (0.79–0.85) | 0.18 (0.15–0.21) |
| 21 | 0.72 (0.24–1.00) | 0.56 (0.40–0.61) | 0.10 (0.09–0.13) | 0.43 (0.40–0.61) | 0.10 (0.09–0.12) | 0.00 (0.00–0.00) | – | – | 0.07 (0.05–0.10) | 0.92 (0.89–0.94) | 0.18 (0.12–0.33) | 0.84 (0.82–0.87) | 0.14 (0.12–0.16) |
| 22 | 0.50 (0.48–0.53) | 0.39 (0.37–0.66) | 0.11 (0.09–0.12) | 0.64 (0.37–0.66) | 0.11 (0.09–0.12) | 0.10 (0.10–0.10) | 0.25 (0.20–0.25) | 0.25 (0.20–0.25) | 0.11 (0.09–0.13) | 0.83 (0.79–0.87) | 0.19 (0.13–0.26) | 0.83 (0.80–0.85) | 0.18 (0.17–0.20) |
| 23 | 0.63 (0.34–0.66) | 0.65 (0.40–0.65) | 0.06 (0.04–0.07) | 0.41 (0.40–0.65) | 0.04 (0.04–0.07) | 0.00 (0.00–0.00) | – | – | 0.09 (0.06–0.11) | 0.92 (0.90–0.94) | 0.39 (0.19–1.65) | 0.84 (0.82–0.87) | 0.15 (0.13–0.17) |
| 24 | 0.69 (0.29–0.72) | 0.70 (0.34–0.71) | 0.08 (0.07–0.08) | 0.35 (0.34–0.71) | 0.07 (0.06–0.08) | 0.00 (0.00–0.00) | 10.0 (10.0–10.78) | 1.00 (1.00–1.00) | 0.19 (0.15–0.22) | 0.84 (0.80–0.87) | 0.37 (0.27–0.51) | 0.75 (0.73–0.78) | 0.22 (0.20–0.24) |
| 25 | 0.65 (0.32–0.69) | 0.67 (0.30–0.67) | 0.08 (0.07–0.09) | 0.31 (0.30–0.67) | 0.08 (0.07–0.09) | 0.30 (0.20–0.30) | 0.20 (0.15–0.20) | 0.20 (0.15–0.20) | 0.08 (0.06–0.10) | 0.82 (0.77–0.87) | 0.22 (0.15–0.56) | 0.84 (0.81–0.86) | 0.14 (0.12–0.16) |
| 26 | 0.53 (0.42–0.59) | 0.62 (0.34–0.64) | 0.10 (0.09–0.11) | 0.36 (0.34–0.64) | 0.10 (0.09–0.11) | 0.00 (0.00–0.00) | – | – | 0.04 (0.03–0.06) | 0.93 (0.90–0.95) | 0.15 (0.06–0.73) | 0.88 (0.86–0.91) | 0.11 (0.09–0.14) |
| 27 | 0.61 (0.36–0.65) | 0.61 (0.31–0.62) | 0.07 (0.06–0.13) | 0.33 (0.31–0.62) | 0.12 (0.07–0.13) | 0.10 (0.10–0.10) | 1.00 (1.00–1.00) | 15.5 (14.0–17.0) | 0.12 (0.10–0.15) | 0.84 (0.80–0.88) | 0.43 (0.31–0.63) | 0.80 (0.77–0.83) | 0.18 (0.16–0.20) |
| 28 | 0.58 (0.38–0.62) | 0.60 (0.27–0.62) | 0.09 (0.06–0.10) | 0.28 (0.26–0.62) | 0.06 (0.06–0.10) | 0.00 (0.00–0.00) | – | – | 0.18 (0.06–0.30) | 0.96 (0.93–0.98) | 0.05 (0.03–0.07) | 0.90 (0.88–0.93) | 0.13 (0.11–0.14) |
| 29 | 0.58 (0.25–0.81) | 0.61 (0.58–0.79) | 0.09 (0.05–0.10) | 0.68 (0.57–0.79) | 0.07 (0.05–0.10) | 0.00 (0.00–0.00) | – | – | 0.11 (0.09–0.15) | 0.87 (0.82–0.91) | 0.31 (0.19–0.58) | 0.83 (0.79–0.85) | 0.18 (0.15–0.20) |
| 31 | 0.61 (0.35–0.65) | 0.69 (0.25–0.70) | 0.08 (0.07–0.09) | 0.26 (0.25–0.70) | 0.08 (0.07–0.09) | 0.30 (0.20–0.30) | 0.15 (0.15–0.15) | 0.15 (0.15–0.15) | 0.04 (0.03–0.06) | 0.92 (0.88–0.95) | 0.06 (0.04–0.12) | 0.89 (0.87–0.92) | 0.12 (0.11–0.14) |
| 32 | 0.53 (0.42–0.58) | 0.67 (0.29–0.68) | 0.08 (0.07–0.09) | 0.30 (0.28–0.68) | 0.08 (0.07–0.09) | 0.00 (0.00–0.00) | 0.30 (0.29–0.35) | 0.30 (0.29–0.35) | 0.09 (0.05–0.18) | 0.93 (0.90–0.95) | 0.18 (0.10–0.56) | 0.87 (0.84–0.89) | 0.13 (0.11–0.16) |
| 33 | 0.51 (0.45–0.55) | 0.63 (0.28–0.65) | 0.11 (0.10–0.12) | 0.29 (0.27–0.65) | 0.11 (0.10–0.12) | 0.10 (0.00–0.10) | 0.25 (0.20–0.25) | 0.25 (0.20–0.25) | 0.65 (0.61–0.68) | 0.48 (0.43–0.53) | 1.13 (0.66–3.05) | 0.51 (0.50–0.53) | 0.50 (0.49–0.51) |
| 34 | 0.52 (0.42–0.59) | 0.64 (0.35–0.66) | 0.10 (0.09–0.11) | 0.37 (0.35–0.66) | 0.10 (0.09–0.11) | 0.00 (0.00–0.00) | – | – | 0.10 (0.07–0.14) | 0.89 (0.86–0.92) | 0.24 (0.14–0.47) | 0.85 (0.83–0.88) | 0.15 (0.13–0.17) |
| 35 | 0.51 (0.44–1.00) | 0.49 (0.39–0.60) | 0.08 (0.06–0.13) | 0.58 (0.39–0.60) | 0.08 (0.08–0.09) | 0.00 (0.00–0.00) | – | – | 0.18 (0.10–0.30) | 0.92 (0.89–0.94) | 0.24 (0.15–0.43) | 0.83 (0.80–0.85) | 0.18 (0.16–0.20) |
| 36 | 0.50 (0.45–0.55) | 0.25 (0.22–0.63) | 0.10 (0.08–0.12) | 0.60 (0.22–0.63) | 0.10 (0.08–0.12) | 0.40 (0.40–0.50) | 0.15 (0.15–0.15) | 0.15 (0.15–0.15) | 0.27 (0.24–0.31) | 0.66 (0.62–0.70) | 1.36 (0.89–2.80) | 0.69 (0.66–0.71) | 0.29 (0.27–0.31) |
| 37 | 1.00 (1.00–1.00) | 0.54 (0.54–0.54) | 0.05 (0.05–0.05) | 0.75 (0.55–0.75) | 0.00 (0.00–0.05) | 0.00 (0.00–0.00) | – | – | 0.06 (0.04–0.09) | 0.96 (0.94–0.97) | 0.16 (0.11–0.23) | 0.85 (0.82–0.87) | 0.13 (0.11–0.15) |
| 38 | 0.83 (0.16–1.00) | 0.54 (0.27–0.55) | 0.09 (0.06–0.13) | 0.28 (0.27–0.54) | 0.07 (0.06–0.09) | 0.10 (0.00–0.10) | 0.25 (0.20–0.30) | 0.25 (0.20–0.30) | 0.26 (0.21–0.37) | 0.76 (0.62–0.80) | 0.58 (0.36–1.34) | 0.70 (0.66–0.73) | 0.25 (0.21–0.30) |
| 39 | 0.58 (0.29–0.75) | 0.43 (0.40–0.74) | 0.11 (0.09–0.13) | 0.68 (0.40–0.76) | 0.11 (0.09–0.13) | 0.00 (0.00–0.10) | 0.30 (0.25–0.35) | 0.30 (0.25–0.35) | 0.11 (0.08–0.25) | 0.87 (0.83–0.90) | 0.19 (0.13–0.29) | 0.83 (0.80–0.86) | 0.17 (0.15–0.20) |
| 41 | 0.50 (0.46–0.53) | 0.62 (0.34–0.63) | 0.11 (0.10–0.12) | 0.35 (0.34–0.63) | 0.11 (0.10–0.13) | 0.10 (0.10–0.10) | 0.30 (0.25–0.35) | 0.30 (0.25–0.35) | 0.19 (0.16–0.22) | 0.70 (0.65–0.75) | 0.39 (0.29–0.51) | 0.73 (0.70–0.76) | 0.23 (0.21–0.25) |
| 42 | 0.67 (0.31–0.70) | 0.71 (0.27–0.72) | 0.09 (0.08–0.10) | 0.28 (0.27–0.72) | 0.09 (0.08–0.10) | 0.40 (0.40–0.40) | 0.25 (0.25–0.25) | 0.25 (0.25–0.25) | 0.12 (0.10–0.15) | 0.73 (0.67–0.78) | 0.32 (0.23–0.50) | 0.81 (0.78–0.83) | 0.17 (0.15–0.19) |
| 43 | 0.60 (0.36–0.65) | 0.64 (0.29–0.66) | 0.11 (0.09–0.12) | 0.31 (0.29–0.66) | 0.11 (0.10–0.12) | 0.40 (0.30–0.40) | 0.15 (0.15–0.20) | 0.15 (0.15–0.20) | 0.14 (0.11–0.16) | 0.67 (0.62–0.72) | 0.73 (0.44–1.28) | 0.79 (0.77–0.82) | 0.20 (0.18–0.22) |
| 44 | 0.63 (0.34–0.67) | 0.61 (0.27–0.62) | 0.09 (0.08–0.13) | 0.29 (0.26–0.62) | 0.12 (0.08–0.13) | 0.10 (0.10–0.20) | 1.00 (1.00–1.00) | 22.0 (20.0–24.0) | 0.08 (0.06–0.10) | 0.82 (0.76–0.88) | 0.19 (0.13–0.30) | 0.82 (0.80–0.85) | 0.14 (0.12–0.16) |
| 45 | 0.50 (0.45–0.55) | 0.64 (0.33–0.65) | 0.09 (0.08–0.11) | 0.35 (0.33–0.65) | 0.09 (0.08–0.11) | 0.20 (0.20–0.20) | 0.20 (0.15–0.20) | 0.20 (0.15–0.20) | 0.12 (0.10–0.14) | 0.77 (0.72–0.82) | 1.15 (0.14–6.49) | 0.82 (0.80–0.85) | 0.19 (0.17–0.22) |
| 46 | 0.62 (0.34–0.67) | 0.71 (0.29–0.72) | 0.10 (0.10–0.13) | 0.31 (0.29–0.72) | 0.12 (0.10–0.13) | 0.30 (0.30–0.40) | 0.30 (0.25–0.30) | 0.30 (0.25–0.30) | 0.15 (0.13–0.18) | 0.46 (0.35–0.56) | 0.30 (0.20–0.48) | 0.78 (0.75–0.80) | 0.21 (0.19–0.23) |
| 48 | 0.61 (0.34–0.67) | 0.68 (0.32–0.71) | 0.11 (0.10–0.14) | 0.34 (0.31–0.71) | 0.11 (0.10–0.14) | 0.00 (0.00–0.00) | 0.30 (0.30–0.30) | 0.30 (0.30–0.30) | 0.11 (0.08–0.13) | 0.74 (0.69–0.79) | 0.24 (0.16–0.34) | 0.81 (0.79–0.84) | 0.16 (0.15–0.18) |
| 49 | 0.72 (0.26–0.74) | 0.57 (0.17–0.58) | 0.08 (0.05–0.09) | 0.18 (0.17–0.58) | 0.06 (0.05–0.09) | 0.20 (0.20–0.20) | 0.20 (0.20–0.25) | 0.20 (0.20–0.25) | 0.22 (0.19–0.24) | 0.79 (0.76–0.83) | 0.59 (0.33–1.52) | 0.78 (0.76–0.80) | 0.27 (0.26–0.29) |

median and 95% bootstrap confidence intervals; D_KL, Kullback-Leibler divergence; Corr, Pearson's correlation coefficient; ChiSq, Chi-Squared statistics; Bhattacharyya, Bhattacharyya distance; $\alpha_{ADE}$ and $\beta_{ADE}$ are grayed out when the median of $w_{ADE}$ is 0.00.

SUPPLEMENTARY TABLE IV
Estimated RP Parameters and Their Evaluation Metrics When Arousal Scale Was Treated As Unimodal

| # | $w_1$ | $\mu_1$ | $\sigma_1$ | $\mu_2$ | $\sigma_2$ | $w_{ADE}$ | $\alpha_{ADE}$ | $\beta_{ADE}$ | $D\_KL$ | Corr | ChiSq | Intersect | Bhattacharyya |
|---|---|---|---|---|---|---|---|---|---|---|---|---|---|
| 11 | 0.53 (0.42–0.59) | 0.61 (0.38–0.62) | 0.07 (0.07–0.08) | 0.39 (0.38–0.62) | 0.07 (0.06–0.08) | 0.00 (0.00–0.00) | – | – | 0.10 (0.07–0.24) | 0.91 (0.88–0.94) | 0.23 (0.15–0.37) | 0.84 (0.80–0.87) | 0.15 (0.13–0.17) |
| 12 | 0.75 (0.21–0.80) | 0.71 (0.28–0.73) | 0.10 (0.08–0.13) | 0.31 (0.28–0.73) | 0.10 (0.08–0.13) | 0.20 (0.20–0.30) | 0.25 (0.20–0.25) | 0.25 (0.20–0.25) | 0.13 (0.11–0.16) | 0.76 (0.68–0.82) | 0.51 (0.31–1.05) | 0.81 (0.78–0.84) | 0.20 (0.18–0.22) |
| 13 | 0.52 (0.38–0.62) | 0.43 (0.41–0.67) | 0.10 (0.08–0.11) | 0.64 (0.41–0.67) | 0.09 (0.08–0.11) | 0.10 (0.10–0.10) | 0.20 (0.15–0.20) | 0.20 (0.15–0.20) | 0.06 (0.04–0.08) | 0.92 (0.88–0.94) | 0.14 (0.08–0.34) | 0.87 (0.84–0.89) | 0.13 (0.11–0.15) |
| 14 | 0.59 (0.37–0.63) | 0.67 (0.33–0.68) | 0.06 (0.06–0.07) | 0.34 (0.33–0.67) | 0.06 (0.06–0.07) | 0.00 (0.00–0.00) | – | – | 0.08 (0.06–0.11) | 0.91 (0.88–0.93) | 0.20 (0.13–0.43) | 0.83 (0.81–0.86) | 0.14 (0.12–0.17) |
| 15 | 0.51 (0.43–0.58) | 0.58 (0.36–0.60) | 0.10 (0.09–0.11) | 0.37 (0.35–0.60) | 0.10 (0.09–0.11) | 0.00 (0.00–0.00) | – | – | 0.04 (0.03–0.06) | 0.95 (0.91–0.97) | 0.06 (0.03–0.11) | 0.89 (0.86–0.92) | 0.12 (0.11–0.14) |
| 16 | 0.53 (0.41–0.59) | 0.64 (0.41–0.65) | 0.08 (0.07–0.11) | 0.42 (0.41–0.65) | 0.09 (0.07–0.11) | 0.00 (0.00–0.00) | – | – | 0.15 (0.11–0.19) | 0.84 (0.79–0.87) | 0.44 (0.31–0.65) | 0.78 (0.75–0.81) | 0.19 (0.17–0.22) |
| 17 | 0.70 (0.20–0.81) | 0.60 (0.35–0.62) | 0.09 (0.08–0.13) | 0.41 (0.34–0.62) | 0.09 (0.08–0.13) | 0.10 (0.10–0.10) | 1.00 (1.00–1.00) | 19.0 (18.0–20.0) | 0.13 (0.10–0.16) | 0.86 (0.81–0.89) | 0.25 (0.17–0.36) | 0.78 (0.75–0.81) | 0.20 (0.18–0.22) |
| 18 | 0.62 (0.36–0.64) | 0.63 (0.24–0.64) | 0.07 (0.07–0.08) | 0.25 (0.24–0.64) | 0.07 (0.07–0.08) | 0.00 (0.00–0.00) | – | – | 0.11 (0.09–0.14) | 0.81 (0.76–0.86) | 0.16 (0.11–0.24) | 0.81 (0.78–0.84) | 0.19 (0.17–0.21) |
| 19 | 0.50 (0.48–0.53) | 0.62 (0.36–0.63) | 0.05 (0.04–0.06) | 0.37 (0.36–0.63) | 0.05 (0.04–0.06) | 0.00 (0.00–0.00) | – | – | 0.13 (0.07–0.23) | 0.95 (0.93–0.97) | 0.36 (0.20–0.93) | 0.86 (0.84–0.88) | 0.15 (0.13–0.18) |
| 21 | 0.72 (0.23–1.00) | 0.55 (0.38–0.60) | 0.10 (0.08–0.13) | 0.41 (0.38–0.61) | 0.10 (0.08–0.11) | 0.00 (0.00–0.00) | – | – | 0.08 (0.06–0.11) | 0.89 (0.86–0.92) | 0.17 (0.12–0.27) | 0.82 (0.79–0.85) | 0.14 (0.12–0.17) |
| 22 | 0.59 (0.37–0.63) | 0.64 (0.37–0.66) | 0.11 (0.10–0.13) | 0.39 (0.37–0.66) | 0.11 (0.10–0.13) | 0.10 (0.10–0.10) | 0.20 (0.20–0.25) | 0.20 (0.20–0.25) | 0.12 (0.10–0.15) | 0.82 (0.77–0.86) | 0.24 (0.16–0.44) | 0.80 (0.77–0.83) | 0.19 (0.18–0.21) |
| 23 | 0.58 (0.38–0.63) | 0.64 (0.39–0.64) | 0.06 (0.04–0.07) | 0.40 (0.39–0.64) | 0.05 (0.04–0.07) | 0.00 (0.00–0.00) | – | – | 0.10 (0.07–0.13) | 0.89 (0.85–0.92) | 0.31 (0.17–1.02) | 0.82 (0.79–0.84) | 0.16 (0.14–0.19) |
| 24 | 0.66 (0.30–0.70) | 0.69 (0.35–0.70) | 0.08 (0.07–0.08) | 0.36 (0.35–0.70) | 0.07 (0.07–0.08) | 0.00 (0.00–0.00) | 10.0 (10.0–10.0) | 1.00 (1.00–1.00) | 0.21 (0.17–0.25) | 0.82 (0.78–0.86) | 0.39 (0.29–0.54) | 0.74 (0.71–0.77) | 0.23 (0.21–0.25) |
| 25 | 0.63 (0.33–0.68) | 0.66 (0.31–0.67) | 0.08 (0.08–0.10) | 0.32 (0.31–0.67) | 0.09 (0.08–0.10) | 0.20 (0.20–0.30) | 0.20 (0.15–0.20) | 0.20 (0.15–0.20) | 0.08 (0.06–0.11) | 0.80 (0.73–0.86) | 0.21 (0.14–0.40) | 0.83 (0.80–0.86) | 0.15 (0.12–0.17) |
| 26 | 0.63 (0.31–0.69) | 0.60 (0.33–0.62) | 0.10 (0.09–0.11) | 0.35 (0.32–0.62) | 0.10 (0.09–0.11) | 0.00 (0.00–0.00) | – | – | 0.05 (0.03–0.06) | 0.93 (0.89–0.96) | 0.29 (0.05–1.25) | 0.89 (0.86–0.91) | 0.12 (0.10–0.14) |
| 27 | 0.68 (0.29–0.72) | 0.62 (0.30–0.63) | 0.08 (0.07–0.13) | 0.32 (0.30–0.63) | 0.12 (0.07–0.13) | 0.10 (0.00–0.10) | 1.00 (1.00–1.00) | 15.0 (14.0–17.0) | 0.12 (0.09–0.15) | 0.85 (0.80–0.89) | 0.47 (0.27–0.72) | 0.80 (0.77–0.83) | 0.18 (0.16–0.20) |
| 28 | 0.65 (0.32–0.68) | 0.61 (0.26–0.63) | 0.09 (0.06–0.10) | 0.27 (0.26–0.63) | 0.07 (0.06–0.10) | 0.00 (0.00–0.00) | – | – | 0.13 (0.05–0.25) | 0.97 (0.94–0.98) | 0.04 (0.03–0.07) | 0.91 (0.89–0.93) | 0.12 (0.11–0.14) |
| 29 | 0.58 (0.25–0.81) | 0.61 (0.58–0.79) | 0.09 (0.05–0.10) | 0.68 (0.57–0.79) | 0.07 (0.05–0.11) | 0.00 (0.00–0.00) | – | – | 0.11 (0.09–0.15) | 0.87 (0.82–0.91) | 0.31 (0.19–0.58) | 0.83 (0.79–0.85) | 0.18 (0.15–0.20) |
| 31 | 0.53 (0.41–0.60) | 0.68 (0.26–0.70) | 0.09 (0.07–0.10) | 0.27 (0.26–0.69) | 0.08 (0.07–0.10) | 0.30 (0.20–0.30) | 0.15 (0.15–0.20) | 0.15 (0.15–0.20) | 0.04 (0.03–0.06) | 0.91 (0.85–0.94) | 0.07 (0.03–0.14) | 0.89 (0.86–0.91) | 0.12 (0.10–0.14) |
| 32 | 0.58 (0.38–0.63) | 0.67 (0.29–0.68) | 0.08 (0.07–0.11) | 0.30 (0.28–0.68) | 0.09 (0.07–0.11) | 0.00 (0.00–0.00) | – | – | 0.07 (0.04–0.11) | 0.91 (0.87–0.94) | 0.16 (0.08–0.43) | 0.86 (0.84–0.89) | 0.13 (0.10–0.15) |
| 33 | 0.55 (0.40–0.61) | 0.63 (0.28–0.66) | 0.11 (0.09–0.14) | 0.30 (0.28–0.66) | 0.11 (0.09–0.14) | 0.00 (0.00–0.00) | 0.25 (0.25–0.25) | 0.25 (0.25–0.25) | 0.61 (0.58–0.66) | 0.53 (0.48–0.58) | 1.30 (0.67–4.35) | 0.54 (0.52–0.55) | 0.49 (0.47–0.50) |
| 34 | 0.51 (0.42–0.59) | 0.62 (0.37–0.64) | 0.10 (0.09–0.12) | 0.39 (0.37–0.64) | 0.11 (0.09–0.12) | 0.00 (0.00–0.00) | – | – | 0.08 (0.06–0.14) | 0.91 (0.87–0.93) | 0.34 (0.19–0.85) | 0.85 (0.82–0.88) | 0.14 (0.12–0.17) |
| 35 | 0.51 (0.42–0.59) | 0.59 (0.40–0.61) | 0.09 (0.07–0.10) | 0.41 (0.39–0.61) | 0.09 (0.07–0.10) | 0.00 (0.00–0.00) | – | – | 0.15 (0.09–0.25) | 0.94 (0.91–0.96) | 0.21 (0.12–0.41) | 0.85 (0.82–0.88) | 0.17 (0.15–0.19) |
| 36 | 0.66 (0.30–0.71) | 0.62 (0.23–0.64) | 0.10 (0.08–0.11) | 0.25 (0.23–0.63) | 0.10 (0.08–0.11) | 0.40 (0.40–0.40) | 0.15 (0.15–0.15) | 0.15 (0.15–0.15) | 0.26 (0.21–0.30) | 0.67 (0.62–0.72) | 0.92 (0.59–1.82) | 0.68 (0.65–0.71) | 0.28 (0.25–0.30) |
| 37 | 0.89 (0.10–1.00) | 0.54 (0.44–0.55) | 0.04 (0.02–0.05) | 0.45 (0.44–0.55) | 0.03 (0.02–0.04) | 0.00 (0.00–0.00) | – | – | 0.08 (0.05–0.16) | 0.95 (0.93–0.97) | 0.12 (0.07–0.25) | 0.85 (0.82–0.87) | 0.12 (0.10–0.15) |
| 38 | 0.85 (0.14–1.00) | 0.54 (0.28–0.55) | 0.09 (0.06–0.15) | 0.28 (0.27–0.55) | 0.07 (0.06–0.15) | 0.10 (0.00–0.10) | 0.25 (0.20–0.25) | 0.25 (0.20–0.25) | 0.27 (0.20–0.38) | 0.74 (0.63–0.80) | 0.56 (0.39–1.17) | 0.69 (0.65–0.73) | 0.26 (0.22–0.30) |
| 39 | 0.62 (0.23–0.81) | 0.41 (0.39–0.74) | 0.12 (0.09–0.13) | 0.66 (0.39–0.77) | 0.12 (0.07–0.13) | 0.00 (0.00–0.10) | 0.30 (0.25–0.30) | 0.30 (0.25–0.30) | 0.13 (0.09–0.18) | 0.87 (0.83–0.91) | 0.33 (0.20–0.66) | 0.82 (0.79–0.85) | 0.19 (0.16–0.22) |
| 41 | 0.50 (0.46–1.00) | 0.50 (0.35–0.63) | 0.12 (0.10–0.18) | 0.37 (0.35–0.63) | 0.12 (0.10–0.13) | 0.10 (0.10–0.10) | 0.30 (0.30–0.35) | 0.30 (0.30–0.35) | 0.17 (0.14–0.21) | 0.75 (0.69–0.79) | 0.34 (0.24–0.49) | 0.74 (0.71–0.77) | 0.22 (0.20–0.25) |
| 42 | 0.73 (0.25–0.76) | 0.73 (0.26–0.74) | 0.09 (0.08–0.10) | 0.28 (0.26–0.74) | 0.10 (0.08–0.10) | 0.40 (0.40–0.40) | 0.25 (0.20–0.25) | 0.25 (0.20–0.25) | 0.08 (0.06–0.12) | 0.82 (0.74–0.87) | 0.18 (0.12–0.29) | 0.83 (0.79–0.86) | 0.14 (0.12–0.17) |
| 43 | 0.73 (0.23–0.77) | 0.66 (0.28–0.67) | 0.10 (0.10–0.12) | 0.30 (0.27–0.67) | 0.11 (0.10–0.12) | 0.40 (0.30–0.40) | 0.20 (0.15–0.20) | 0.20 (0.15–0.20) | 0.10 (0.08–0.13) | 0.75 (0.69–0.80) | 0.36 (0.22–0.66) | 0.82 (0.79–0.85) | 0.16 (0.14–0.19) |
| 44 | 0.72 (0.26–0.75) | 0.62 (0.26–0.62) | 0.08 (0.07–0.12) | 0.28 (0.26–0.62) | 0.12 (0.07–0.13) | 0.10 (0.10–0.10) | 1.00 (1.00–1.00) | 22.0 (20.0–24.02) | 0.06 (0.05–0.09) | 0.91 (0.87–0.94) | 0.15 (0.10–0.28) | 0.85 (0.82–0.88) | 0.13 (0.11–0.15) |
| 45 | 0.50 (0.46–0.54) | 0.64 (0.32–0.65) | 0.08 (0.07–0.11) | 0.34 (0.32–0.65) | 0.10 (0.07–0.11) | 0.20 (0.10–0.20) | 0.20 (0.20–0.20) | 0.20 (0.20–0.20) | 0.13 (0.11–0.16) | 0.77 (0.71–0.82) | 2.85 (0.15–13.1) | 0.81 (0.78–0.83) | 0.21 (0.18–0.23) |
| 46 | 0.58 (0.38–0.63) | 0.71 (0.30–0.72) | 0.11 (0.10–0.13) | 0.32 (0.30–0.72) | 0.12 (0.10–0.13) | 0.40 (0.30–0.50) | 0.25 (0.25–0.30) | 0.25 (0.25–0.30) | 0.14 (0.11–0.17) | 0.52 (0.39–0.63) | 0.25 (0.16–0.40) | 0.79 (0.76–0.82) | 0.20 (0.18–0.22) |
| 48 | 0.52 (0.42–0.59) | 0.67 (0.31–0.69) | 0.11 (0.08–0.12) | 0.34 (0.31–0.69) | 0.10 (0.09–0.12) | 0.00 (0.00–0.00) | – | – | 0.10 (0.07–0.13) | 0.75 (0.70–0.81) | 0.20 (0.14–0.31) | 0.82 (0.79–0.85) | 0.16 (0.14–0.18) |
| 49 | 0.78 (0.21–0.80) | 0.58 (0.17–0.59) | 0.07 (0.05–0.08) | 0.18 (0.17–0.59) | 0.06 (0.05–0.08) | 0.20 (0.20–0.20) | 0.20 (0.20–0.25) | 0.20 (0.20–0.25) | 0.18 (0.16–0.22) | 0.87 (0.83–0.90) | 0.34 (0.19–0.93) | 0.79 (0.77–0.81) | 0.26 (0.24–0.28) |

median and 95% bootstrap confidence intervals; D_KL, Kullback-Leibler divergence; Corr, Pearson's correlation coefficient; ChiSq, Chi-Squared statistics; Bhattacharyya, Bhattacharyya distance; $\alpha_{ADE}$ and $\beta_{ADE}$ are grayed out when the median of $w_{ADE}$ is 0.00.